\documentclass{JHEP3}    
\usepackage{epsfig,amsfonts,amsmath}

\newcommand{\eref}[1]{(\ref{#1})}

\newcommand{\fref}[1]{Figure~\ref{#1}}
\newcommand{\cref}[1]{Chapter~\ref{#1}}
\newcommand{\beq}{\begin{equation}}
\newcommand{\eeq}{\end{equation}}
\newcommand{\ba}{\begin{array}}
\newcommand{\ea}{\end{array}}
\newcommand{\bcenter}{\begin{center}}
\newcommand{\ecenter}{\end{center}}

\def\C{\mathbb{C}}
\def\IC{\mathbb{C}}

\def\IGa{\relax\hbox{${\rm I}\kern-.18em\Gamma$}}

\def\IP{\mathbb{P}}

\def\Z{\mathbb{Z}}
\def\IZ{\mathbb{Z}}
\def\II{\mathbb{I}}

\def\N{\mathcal{N}}

\def\codim{{\mathop{\rm codim}}}

\def\rank{{\rm rank}}

\def\Diff{{\rm Diff}}

\def\Hom{{\rm Hom}}
\def\dim{{\rm dim}}

\def\odd{{\rm odd}}
\def\even{{\rm even}}

\def\smiley{\hbox{\large$\bigcirc$\hspace{-0.80em}\raise.2ex
\hbox{$\cdot\cdot$}\kern-.61em\lower.2ex\hbox{\scriptsize$\smile$}}\ }
\def\frowny{\hbox{\large$\bigcirc$\hspace{-0.80em}\raise.2ex
\hbox{$\cdot\cdot$}\kern-.635em\lower.2ex\hbox{\scriptsize$\frown$}}\ }

\makeatletter
\let\hangafter\@hangfrom
\makeatother

\newcommand{\be}{\begin{equation}}
\newcommand{\ee}{\end{equation}}
\newcommand{\bea}{\begin{eqnarray}}
\newcommand{\eea}{\end{eqnarray}}
\newcommand{\bean}{\begin{eqnarray*}}
\newcommand{\eean}{\end{eqnarray*}}
\newcommand{\bc}{\begin{center}}
\newcommand{\ec}{\end{center}}

\preprint{MIT-CTP-3619\\ {\tt hep-th/0504110}}

\title{Brane Dimers and Quiver Gauge Theories}

\author{Sebasti\'an Franco$^1$, Amihay Hanany$^1$, Kristian D. Kennaway$^2$, David Vegh$^1$, Brian Wecht$^1$
\\
~\\
$^1$ Center for Theoretical Physics,
Massachusetts Institute of Technology,\\
Cambridge, MA 02139, USA.\footnote{
Research supported in part by the CTP and the LNS
of MIT and the U.S. Department of Energy under cooperative agreement
$\#$DE-FC02-94ER40818. AH is also supported in part by the BSF American--Israeli Bi--National Science Foundation and 
a DOE OJI award. BW is supported in part by National Science Foundation Grant 
beas
PHY-00-96515.
DV is supported by the MIT Praecis Presidential Fellowship.  }\\~\\

$^2$ Department of Physics, 
University of Toronto,\\
Toronto, ON M5S 1A7, CANADA.\footnote{K.K. is supported by NSERC.}
}

\abstract{We describe a technique which enables one to quickly compute an infinite number of toric
geometries and their dual quiver gauge theories. The central object in this
construction is a ``brane tiling,'' which is a collection of D5-branes ending on an NS5-brane
wrapping a holomorphic curve that
can be represented as a periodic tiling of the plane. 
This construction solves the longstanding problem of computing superpotentials for D-branes probing a singular non-compact toric Calabi-Yau manifold, and overcomes many difficulties which were encountered in previous work. The brane tilings give the largest class of $\N=1$ quiver gauge theories yet studied. A central feature of this work is the relation of these tilings to dimer constructions 
previously studied in a variety of contexts.
We do many examples of computations with dimers, which give new results
as well as confirm previous computations. Using our methods we explicitly
derive the moduli space of the entire $Y^{p,q}$ family of quiver
theories, verifying that they correspond to the appropriate
geometries. Our results may be interpreted as a generalization of the McKay correspondence
to non-compact 3-dimensional toric Calabi-Yau manifolds.}

\begin{document}

\section{Introduction}
Shortly after the discovery of the importance of D-branes in string theory, it became clear that they provide a deep connection between algebra and geometry. This is realized in string theory in the following way: the D-brane, a physical object in spacetime, probes the geometry in which it lives, and the properties of spacetime fields are reflected in its worldvolume gauge theory. On the other hand, the D-brane has fundamental strings ending on it, thus giving rise to gauge quantum numbers which enumerate the possible ways strings can end on the brane when it is embedded in the given singular geometry. This fact leads to a pattern of algebraic relations, which are conveniently encoded in terms of algebraic quantities like Dynkin diagrams and, more generally, quiver gauge theories.

One simple example is given by a collection of $N$ parallel D-branes; these have fundamental strings stretching between them and are in one-to-one correspondence with Dynkin diagrams of type $A_{N-1}$. Branes are mapped to nodes in the Dynkin diagrams and fundamental strings are mapped to lines. Placing this configuration on a circle leads to an affine Dynkin diagram $\hat A_{N-1}$, where the imaginary root is mapped to a fundamental string encircling the compact direction. Many more examples of this type lead to a beautiful relationship between branes and Lie algebras.

When D-branes in Type II string theory are placed on an ALE singularity of ADE type, the gauge theory living on them is encoded by an affine ADE Dynkin diagram. Here, fractional branes are mapped to nodes in the Dynkin diagram while strings stretching between the fractional branes are mapped to lines in the Dynkin diagram. The gauge theory living on the branes has 8 supercharges, since the ALE singularity breaks one half of the supersymmetry while the D-branes break a further half. With this amount of supersymmetry ($\N=2$ in four dimensions), it is enough to specify the gauge group and the matter content in order to fix the Lagrangian of the theory uniquely. As above, we see that this gauge theory also realizes the relationship between algebra and geometry: on the algebra side, we have the Dynkin diagram, and on the geometry side there is a moduli space of vacua which is the ADE type singularity. This connection between the Lie algebras of affine Dynkin diagrams and the geometry of ALE spaces is known as the McKay correspondence \cite{mckay}. This relation was first derived in the mathematics literature and later -- with the help of D-branes -- became an important relation in string theory.

The McKay correspondence can be realized in other ways in string theory. One example which will be important to us in this paper is the configuration of NS5-branes and D-branes stretching between them which was studied in \cite{Hanany:1996ie}. A collection of $N$ NS5-branes with D-branes stretching between them results in a  gauge theory on the D-branes which turns out to be encoded by an $A_{N-1}$ Dynkin diagram. Here, NS5-branes are mapped to lines while D-branes stretching between NS5-branes are mapped to nodes in the quiver gauge theory. Putting this configuration on a circle leads to an affine version of this correspondence $\hat A_{N-1}$; the imaginary root now corresponds to a D-brane which wraps the circle. This correspondence is very similar to the D-brane picture which was presented above and indeed, using a chain of S- and T- dualities, one can get from one configuration of branes to the other, while keeping the algebraic structure the same. Furthermore, the gauge theory living on a D-brane stretched between NS5-branes is very similar to the gauge theory living on a D-brane that probes an $A_{N-1}$ singularity. Indeed, as first observed in \cite{Ooguri:1995wj}, the collection of $N$ NS5-branes on a circle is T-dual to an ALE singularity of type $A_{N-1}$ with one circular direction. A detailed study of this correspondence was performed in \cite{Karch:1998yv}.

Many attempts at generalizing the McKay correspondence from a 2 complex dimensional space to a 3 complex dimensional space have been made in the literature
\cite{Ito:1994zx,Reid:1997zy,Govindarajan:2000vi,Tomasiello:2000ym,Mayr:2000as,Lerche:2001vj}. From the point of view of branes in string theory it is natural to extend the 2-dimensional correspondence stated above from D-branes probing a 2-dimensional singular manifold to a collection of D-branes probing a 3 dimensional singular Calabi-Yau (CY) manifold
\cite{Douglas:1997de}.  A few qualitative features are different in this case. First, the supersymmetry of the gauge theory living on the D-brane is now reduced to 4 supercharges. This implies that gauge fields and matter fields are not enough to uniquely determine the Lagrangian of the theory, and one must also specify a superpotential which encodes the interactions between the matter fields.

This is an important observation: any attempt at stating the 3-dimensional McKay correspondence must incorporate the superpotential, which was uniquely constrained in the 2-dimensional case. Second, the matter multiplets in theories with 4 supercharges are chiral and therefore have a natural orientation. In the theories with 8 supercharges, for every chiral multiplet there is another chiral multiplet with an opposite orientation, transforming together in a hypermultiplet. Therefore, an overall orientation is not present in a theory with 8 supercharges. We conclude that the 3-dimensional McKay correspondence requires information about these orientations, absent in the 2-dimensional case.

Studying the first few examples for the 3-dimensional McKay correspondence (the simplest of which is the orbifold $\C^3/\Z_3$), it became clear that the objects which replace the Dynkin diagrams are quivers with oriented arrows \cite{Douglas:1996sw}. For these objects, nodes represent gauge groups, oriented arrows between two nodes represent bifundamental chiral multiplets, and certain closed paths in the quiver (which represent gauge-invariant operators) represent terms in the superpotential.  It is important to note that only a subset of all closed paths on the quiver appears in the superpotential, and finding which particular subset is selected for the quiver associated to a given toric singularity is a difficult task. This difficulty will be greatly simplified with the results of this paper. 

Since we will be using quivers throughout this work, it will be useful to briefly recall what is 
currently known about the theories we can study via string theory. The first known examples of 
quiver theories obtained from string theory were those dual to $\IC^3/\Gamma$, where $\Gamma$ is any 
discrete subgroup of $SU(3)$ \cite{Lawrence:1998ja,Hanany:1998sd}. The most common examples of this type take
$\Gamma = \IZ_n$ or $\Gamma=\IZ_n \times \IZ_m$. These theories are easy to construct, 
since it is straightforward to write down an orbifold action on the coordinates of
$\IC^3$. If $| \Gamma | = k$, then there are $k$ nodes in the dual quiver and a bifundamental
for each orbifold action on $\IC^3$ that connects different regions of the covering space.
These orbifold theories may be described torically in a straightforward manner. It was
then realized that partial resolution of these orbifold spaces corresponds to 
Higgsing the quivers; in this manner, people were able to obtain many different quiver
theories and their dual toric geometries \cite{Feng:2000mi,Feng:2001xr,Beasley:2001zp}. 
For a good review of toric geometry, see \cite{Fulton,Leung:1997tw}.

It was not long before a general algorithm for deriving toric data from a given quiver was 
found; this procedure is usually called ``The Forward Algorithm'' \cite{Feng:2000mi}. Although the procedure
is well-understood, it is computationally prohibitive for quivers with more than approximately ten nodes.
The Forward Algorithm, in addition to providing the toric data for a given quiver theory, also
gives the relative {\bf multiplicities} of the gauged linear sigma model (GLSM) fields in the related sigma model \cite{Feng:2002zw}. However, the 
same problem applies here as well: the toric diagrams and their associated multiplicities are difficult 
to derive for large quivers.

In recent months there has been much progress in the arena of gauge theories dual to
toric geometries. Gauntlett, Martelli, Sparks, and Waldram \cite{Gauntlett:2004yd} found an infinite class of
Sasaki-Einstein (SE) metrics; previous to their work, only two explicit SE metrics were known.
These metrics are denoted $Y^{p,q}$ and depend only on two integers $p$ and $q$,
where $0 < q < p$. 
In related work, Martelli and Sparks \cite{Martelli:2004wu} found the toric descriptions of the $Y^{p,q}$ theories,
and noted that some of these spaces were already familiar, although their
metrics had not previously been known. One of the simplest examples is
$Y^{2,1}$, which turns out to be the SE manifold which is the base of the complex cone over
the first del Pezzo surface.  The R-charges
for $Y^{2,1}$ were computed in \cite{Bertolini:2004xf} and shown to agree exactly with the geometrical
computation done using the metric found in \cite{Martelli:2004wu}.  More progress was made when the gauge theory duals of the $Y^{p,q}$ spaces were found
\cite{Benvenuti:2004dy},
providing an infinite class of AdS/CFT dual pairs. 
These theories have survived
many nontrivial checks of the AdS/CFT correspondence, such as central charge and R-charge computations 
from volume calculations on the string side and $a$-maximization \cite{Intriligator:2003jj} on the gauge theory side. Inspired by these
gauge theories, there have since been many new and startling checks of AdS/CFT, such as the construction of 
gravity duals for cascading RG flows \cite{Herzog:2004tr}. Thus, there
has been remarkable progress recently in the study of toric Calabi-Yau manifolds and
their dual gauge theories; however, a general procedure for constructing
the dual to a given CY is still unknown. In this work, we will shed some light on this
problem. 

One of the results of the present work is that the ingredients required to uniquely define an $\mathcal{N}=1$ quiver gauge theory -- gauge groups, chiral matter fields, and superpotential terms -- may be represented in terms of nodes, lines and faces of {\it a single object}, which is the quiver redrawn as a planar graph on the torus (for the quiver theories corresponding to toric singularities). This point will be crucial in the construction of the quiver gauge theory using dimers, as will be discussed in detail in section \ref{sec:dimers}.

One may also ask how these theories may be constructed in string theory by using branes, as explained above for the case of theories with 8 supercharges. A key observation is that if a collection of $m$ NS5-branes is T-dual to an orbifold $\C^3/\IZ_m$, then a collection of $m$ NS5-branes intersecting with $n$ NS$5^\prime$-branes with both sets of NS5-branes sharing 3+1 space-time directions, is equivalent under two  T-dualities to an orbifold singularity of type $\C^3/(\IZ_m\times\IZ_n)$. When D3-brane probes are added over the orbifold, they are mapped to D5-branes suspended between the NS5-branes on the T-dual
configuration. Indeed, a study of these theories using the Brane Box Models of \cite{Hanany:1997tb} was done in \cite{Hanany:1998it}. Another important development in the brane construction of quiver gauge theories with 4 supercharges was made in \cite{Aganagic:1999fe} where it was realized that the quiver gauge theories which live on D-branes probing the conifold and its various orbifolds are constructed by ``Brane Diamonds.'' Brane diamonds were
also applied to the study of gauge theories for D-branes probing complex cones over del Pezzo surfaces \cite{Feng:2001bn}.

In the present paper, we consider a more intricate configuration of branes. First, we take
an NS5-brane which extends in the 0123 directions and wraps a complex curve $f(x,y)=0$,
where $x$ and $y$ are holomorphic coordinates in the 45 and 67 directions, respectively. 
We typically depict this by drawing this curve in the 4 and 6 directions, where
it looks like a network that separates the plane into different regions, i.e.~a tiling\footnote{Related work, on how to tile a domain wall with lower-dimensional domain walls, was done in \cite{Bazeia:2000mx}.}.
We do not explicitly write down the equation for this curve, but do note that
a requirement of our construction is that the tiling of the 46 plane is such that
all polygons have an even number of sides.
The 4 and 6 directions are compact, forming a torus, and we take the D5 branes to be finite in these directions (but extended in the 0123 directions) and bounded by the curve which is wrapped by the NS5-brane. As above, this brane configuration results in a quiver gauge theory living on the D5-branes. 
The rules for computing this quiver theory turn out to follow similar guidelines to those in the constructions mentioned above: gauge groups are faces of the intersecting brane configuration, bifundamental fields arise across NS5-branes which are lines in the brane configuration, and superpotential terms show up as vertices.

It is important to note that the Brane Box Models were formulated using periodic square graphs for encoding the rules of the quiver gauge theory, but it will become clear in this paper that the correct objects to use to recover that construction are hexagonal graphs, and in fact the brane boxes are recovered in a degenerate limit in which two opposite edges of the hexagons are reduced to zero length.

We observe that in both the brane box and diamond constructions, the brane configurations are related to the quiver gauge theory in the following sense: faces in the brane configuration are mapped to nodes in the quiver, lines are mapped to orthogonal lines and nodes are mapped to faces. The statement of this duality will be formulated precisely in Section \ref{sec:unity}, and will prove to be a very powerful tool in generalizing these constructions to a larger class of quiver gauge theories (those whose moduli space describe non-compact toric CY 3-folds).

Thus, we find that it is possible to encode {\bf all the data} necessary to uniquely
specify an $\mathcal{N}=1$ quiver gauge theory in a tiling of the plane. The dual graph is then essentially the
quiver theory, written in such a way as to encode the superpotential data as well. 
As we will now see, however, this tiling encodes much more than just the quiver theory -- 
it also encodes the dual toric geometry! The central object for deriving the toric geometry
is the {\bf dimer}, which we now explain.

Since we have taken our brane tiling to consist of polygons with an even number
of sides, and all cycles of our periodic graph have even length, it is always possible to color the nodes of the graph with two different colors
(say, black and white) in such a way that any given black node is adjacent only to white nodes, and vice
versa. Such graphs are well-known in condensed matter physics, where the links between
black nodes and white nodes are called dimers; one may think of a substance formed out
of two different type of atoms (e.g.~a salt), where a dimer is just an edge of
the lattice with a different atom at each end. One can allow bonds between adjacent atoms
to break and then re-form in a possibly different configuration; the statistical mechanics
of such systems has been extensively studied. 

Recently, dimers have shown up in the context of string theory on toric Calabi-Yau manifolds. 
In \cite{Okounkov:2003sp},
the authors propose a relationship between the statistical mechanics of dimer models
and topological strings on a toric non-compact Calabi-Yau. 
The relationship between toric geometry and dimer models was developed further in \cite{Hanany:2005ve}, where it was
shown how it is possible to obtain toric diagrams and GLSM multiplicities via dimer techniques. In general, however, we expect
that one should be able to derive the quiver gauge theory dual to any given toric geometry. This is the purpose of the present work,
to describe how dimer technology may be used to efficiently derive both the quiver theory and the toric geometry, thus
giving a fast and straightforward way of deriving AdS/CFT dual pairs. 

We can now state that the 3-dimensional McKay correspondence is represented in string theory as a physical brane configuration of an
NS5-brane spanning four dimensions and wrapping an holomorphic curve on four other dimensions, and D5-branes. Alternatively, we can use a twice T-dual (along the 4 and 6 directions) description: the McKay correspondence is realized by the quiver gauge theory that lives on D-branes probing toric CY 3-folds.
As a byproduct of these two equivalent representations we can argue that it is possible to find NS5-brane configurations that are twice T-dual to these toric singular CY manifolds. One removes the D-branes and ends up with NS5-branes on one side and singular geometries on the other. 

The outline of this paper is as follows. In Section \ref{section_tilings}, we summarize the
basic features of our construction, and establish the relationship between
brane tilings and quiver gauge theories. We explain the brane construction that
leads to the quiver theory, and detail how it is possible
to read off all relevant data about the quiver theory from the brane tiling. 
We derive an interesting identity for which the brane tiling perspective 
provides a simple proof.  We illustrate
these ideas with a simple example, Model I of del Pezzo 3. Additionally, we describe
a new object, the ``periodic quiver," which is the dual graph to the brane tiling
and neatly summarizes the quiver and superpotential data for a given gauge theory. 

In Section \ref{sec:dimers}, we describe the utility of the dimer model and
review the relationship between dimers and toric geometries. We begin
by summarizing relevant facts about dimer models which we will
use repeatedly throughout the paper. The
central object in any computation is the Kasteleyn matrix, which is a weighted 
adjacency matrix that is easy to derive. We do a simple computation as an example, which
illustrates the basic techniques required to compute the toric geometry related
to any given brane tiling. 

Section \ref{section_dimers_GLSM} provides the relationship between dimers
and fields in the related gauged linear sigma model. We review the relationship of
toric geometries to GLSMs, and describe how the dimer model allows one to compute
multiplicities of GLSM. These techniques are illustrated with an example, that of
the Suspended Pinch Point (SPP) \cite{Morrison:1998cs}.

Section \ref{section_massive_nodes} briefly describes how massive
fields arise via the brane tiling description, and comments on the
process of integrating out these fields from the perspective of both the
brane tiling and the Kasteleyn matrix. Section \ref{sec:seiberg} 
talks about Seiberg duality from three complementary perspectives: 
the brane tiling, the quiver, and the Kasteleyn matrix. We illustrate these
viewpoints with ${\bf F}_0$ as an example. 

Section  \ref{section_partial_resolution} gives two descriptions
of the process of partial resolution of orbifold singularities, both 
from the brane tiling and quiver perspectives. In Section \ref{section_different_superpotentials},
we describe how one may construct brane tilings which produce identical quivers
but different superpotentials; this is illustrated via the quiver from Model II of ${\bf dP}_3$.
In Section \ref{sec:examples}, we present many different examples of brane tilings,
and compute the Kasteleyn matrix and dual toric geometry in each example. These computations
duplicate known results, as well as generate new ones. Most notably, we
find that toric diagrams with specified GLSM multiplicities are not in
one-to-one correspondence with toric phases of quiver gauge theories, as had
previously been suspected. This computation is done for Pseudo del Pezzo 5, where
we find two toric phases with identical GLSM multiplicities.
Finally, in Section \ref{section_conclusions}, we briefly conclude
and present some suggestions for further study. 

\section{Brane tilings and quivers}
\label{section_tilings}

In this section we introduce the concept of {\bf brane tilings}. They are Type IIB configurations of NS5 and D5-branes that generalize the brane box \cite{Hanany:1998it} and brane diamond \cite{Aganagic:1999fe} constructions and are dual to gauge theories on D3-branes transverse to arbitrary {\bf toric} singularities. From now on, we proceed assuming that the dual geometry is toric and introduce the relevant brane configurations. The reason for the requirement that the corresponding singularities are toric will become clear in this and subsequent sections.

In our construction, the NS5-brane extends in the 0123 directions and wraps a holomorphic curve embedded in the 4567 directions (the 46 directions are taken to be compact). D5-branes span the 012346 directions and stretch inside the holes in the NS5 skeleton like soap bubbles. The D5-branes are bounded by NS5-branes in the 46 directions, leading to a 3+1 dimensional theory in their world-volume at low energies. The branes break supersymmetry to 1/8 of the original value, leading to 4 supercharges, i.e.~ $\mathcal{N}=1$ in four dimensions. In principle, there can be a different number of D5-branes $N_I$ in each stack. This would lead to a product gauge group $\prod_I SU(N_I)$.  Strings stretching between D5-branes in a given stack give rise to the gauge bosons of $SU(N_I)$ while strings connecting D5-branes in adjacent stacks $I$ and $J$ correspond to states in the bifundamental of $SU(N_I)\times SU(N_J)$. 
We will restrict ourselves to the case $N_I=N$ for all $I$. Theories satisfying this restriction on the ranks were dubbed {\bf toric phases}
in \cite{Feng:2002zw}, We should emphasize though, that there are quivers that are dual to toric geometries but that do not satisfy this condition.

It is worthwhile here to note a few properties of NS5-branes that are relevant for this construction. As is well-known, an
NS5-brane backreacts on its surrounding spacetime to create a throat geometry. When we have two sets of D5-branes ending
on different sides of the NS5-brane, the throat separates the two sets of branes. The D-branes may then
only interact via fundamental strings stretching between them; these are the bifundamentals in the quiver gauge theory.
Initially it might seem like there are two conjugate bifundamentals which pair up to form hypermultiplets, but
in this case, where the NS5-brane wraps a holomorphic curve, the orientation of the NS5-brane projects one of these out of the massless spectrum \cite{Elitzur:2001zh}. Thus the resulting
quiver theory will generically have arrows pointing in only one direction (it is easy to get quivers with bidirectional
arrows as well, but these will instead come from strings stretching across different NS5-branes rather than
both orientations across the same NS5-brane).

The important physics is captured by drawing the brane tiling in the 46 plane. The NS5-branes wrap a holomorphic curve, the real section of which is a graph $G$ in the 46 plane, which we will later show must be $\bf bipartite$.  A graph is bipartite when its nodes can be colored in white and black, such that edges only connect black nodes to white nodes and vice versa. By construction, $G$ is $\IZ^2$-periodic under translations in the 46 plane since these directions are taken to be compact. We will see in the next section that the existence of $G$ is associated to the duality between quiver gauge theories and  dimer models.

Given a brane tiling, it is straightforward to derive its associated quiver gauge theory. The brane tiling encodes both the quiver diagram and the superpotential, which can be constructed according to the dictionary given in Table \ref{table:dict} (see the following section).  Conversely, we can use this set of rules to construct a brane tiling from a given quiver with a superpotential. In the following section we will make this correspondence precise.

\begin{table}[th]
\begin{tabular}{l|l|l}
\hline
{\bf Brane tiling} & {\bf String theory} & {\bf Gauge theory} \\ \hline \hline 
$2n$-sided face & D5-branes & Gauge group with $n$ flavors \\ \hline
Edge between two & String stretched between D5-  & Bifundamental chiral multiplet\\
 polygons I and J               &branes through NS5 brane. &  between gauge groups I and J; \\ 
 & & We orient the arrow such that \\
 & & the white node is to the right. \\ \hline 
k-valent vertex & Region where $k$ strings & Interaction between $k$ chiral \\
 & interact locally.&multiplets, i.e.~order $k$ term in\\
 && the superpotential. The signs for \\
      &         & the superpotential terms are\\
      && assigned such that white and \\
         &      & black nodes correspond to plus \\
&&         and minus signs respectively. \\ \hline
\end{tabular}
\caption{Dictionary for translating between brane tiling, string theory and gauge theory objects.\label{table:dict}}
\end{table}

Several interesting consequences follow naturally from this simple set of rules.
Some of them are well known, while others are new. The fact that the graphs under consideration are bipartite implies that each edge has a black and a white endpoint. Edges correspond to bifundamental fields while nodes indicate superpotential terms, with their sign determined by the color of the node. Thus, we conclude that each bifundamental field appears exactly twice in the superpotential, once with a plus and once with a minus sign. We refer to this as the {\bf toric condition} and it follows from the underlying geometry being an affine toric variety \cite{Feng:2002zw}.

The total number of nodes inside a unit cell is even (there are equal numbers of black and white nodes). Thus, we conclude that the total number of terms in the superpotential of a quiver theory for a toric singularity is even. Although this condition is reminiscent of the toric condition, it is different. It is comforting to see that it is satisfied by all the examples in the literature (orbifolds, del Pezzos, ${\bf F}_0$, pseudo-del Pezzos, SPP, $Y^{p,q}$, $X^{p,q}$, etc).

Bidirectional arrows and even adjoint fields in the quiver can be simply implemented in this construction, by suitably choosing the adjacency of polygons. We will present an example containing both situations in section \ref{sec:spp}.

Let us define
\beq
\begin{array}{lc|cl}
\ \ \ \ \mbox{{\bf Brane tiling}}   & \ \ & \ \ &\ \ \ \ \ \ \ \ \ \ \mbox{{\bf Gauge theory}} \\ \hline \hline
F \mbox{: number of faces} & & & N_g \mbox{: number of gauge groups} \\
E \mbox{: number of edges} & & & N_f \mbox{: number of fields} \\
N \mbox{: number of nodes} & & & N_W \mbox{: number of superpotential terms} 
\end{array}
\nonumber
\label{dictionary}
\eeq
According to the dictionary above, $F=N_g$, $E=N_f$ and $N=N_W$. Applying Euler's formula to a unit cell in the graph, we see that $F+N-E=2g-2=0$ (where we have used that the graph lives on the torus), which translates into the following identity for quiver theories\footnote{This identity was derived empirically with Barak Kol using the known examples. The brane tiling gives a proof
for a generic $\N=1$ toric theory.}:

\beq
N_g+N_W-N_f=0.
\label{Euler_quiver}
\eeq
The geometric intuition we gain when using brane tilings make the derivation of this remarkable identity straightforward.

It is interesting to point out here that the Euler formula has another interpretation. Let us
assign an R-charge to each bifundamental field in the quiver, i.e. to each
edge in the brane tiling. At the IR superconformal fixed point, we
know that each term in the superpotential must satisfy
\beq
\sum_{i \in edges\,\, around \,\, node} R_i =2 \qquad {\rm for\,\, each \,\,node}
\label{wcharge}
\eeq
where the sum is over all edges surrounding a given node.
We can sum over all the nodes in the tiling, each of which corresponds to a superpotential
term, to get $\sum_{edges,nodes} R = 2N$. Additionally, the beta function for each
gauge coupling must vanish,
\beq
2 + \sum_{i \in edges\,\, around \,\, face} (R_i-1) =0  \qquad {\rm for\,\, each \,\,face}
\eeq
where the sum is over all edges surrounding a given face. 
But we can now sum this over all the faces in the tiling to get
$2F + 2N - 2E = 0$, where we have used the fact that the double sum hits
every edge twice, and (\ref{wcharge}). The sums $\sum_{edges, nodes} R$ 
and $\sum_{edges, faces} R$ are equal because each double sum
has the R-charge of each bifundamental contributing twice. 
Thus we see that the requirements that
the superpotential have $R(W)=2$ and the beta functions vanish (i.e. that the
theory is superconformal in the IR) imply that
the Euler characteristic of the tiling is zero. 
This condition is
the analog of a similar condition for superconformal quivers discussed in \cite{Intriligator:2003wr,Benvenuti:2004dw}.
Conversely we see that, in the case in which the ranks of all gauge groups are equal,
the construction of tilings over Riemann surfaces different from a torus leads to non-conformal
gauge theories.

\begin{figure}[ht]
  \epsfxsize = 15cm
  \centerline{\epsfbox{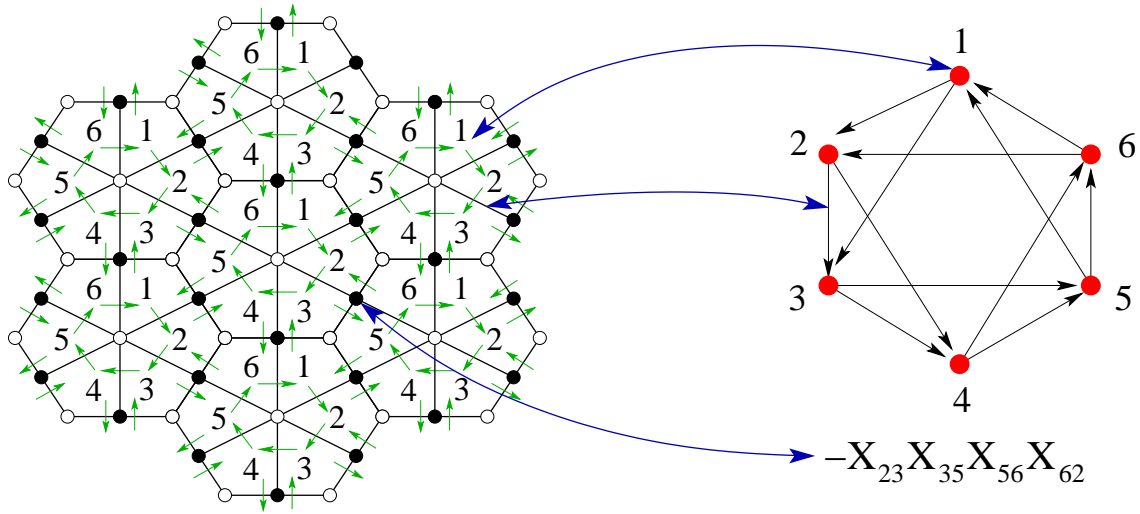}}
  \caption{A finite region in the infinite brane tiling and quiver diagram for Model I of ${\bf dP}_3$. We indicate the correspondence between: gauge groups $\leftrightarrow$ faces, bifundamental fields $\leftrightarrow$ edges and superpotential terms $\leftrightarrow$ nodes.}
  \label{dimer_quiver_dP3_1}
\end{figure}

Let us illustrate the concepts introduced in this section with a simple example, one of the toric phases of ${\bf dP}_3$, denoted Model I in \cite{Feng:2002zw}. Its corresponding quiver diagram is presented in \fref{dimer_quiver_dP3_1} and its superpotential is 

\beq
\begin{array}{ll}
W &=X_{12}X_{23}X_{34}X_{45}X_{56}X_{61}-(X_{23}X_{35}X_{56}X_{62}+X_{13}X_{34}X_{46}X_{61}+X_{12}X_{24}X_{45}X_{51}) \\ 
  &+(X_{13}X_{35}X_{51}+X_{24}X_{46}X_{62}).
\end{array}
\label{W_dP3_1}
\eeq
The quiver diagram has 6 gauge groups and 12 bifundamental fields. Hence, the brane configuration will have 6 faces and 12 edges in a unit cell. The superpotential \eref{W_dP3_1} has 1 order six, 3 quartic and 2 cubic terms. According to \eref{dictionary} we thus have 1 6-valent, 3 4-valent and 2 3-valent nodes. The final brane tiling is shown in \fref{dimer_quiver_dP3_1}.

\subsection{Unification of quiver and superpotential data}
\label{sec:unity}

An $\mathcal{N}=1$ quiver gauge theory is described by the following data: a directed graph representing the gauge groups and matter content, and a set of closed paths on the graph representing the gauge invariant interactions in the superpotential.  
An equivalent way to characterise this data is to view it as defining a CW-complex; in other words, we may take the superpotential terms to define the 2-dimensional faces of the complex bounded by a given set of edges and vertices (the 1-skeleton and 0-skeleton of the complex).  Thus, the quiver and superpotential may be combined into a single object, a planar tiling of a 2-dimensional (possibly singular) space.  Toric quiver theories, as we will see, are defined by planar tilings of the 2-dimensional torus.

\begin{figure}[ht]
  \epsfxsize = 15cm
  \centerline{\epsfbox{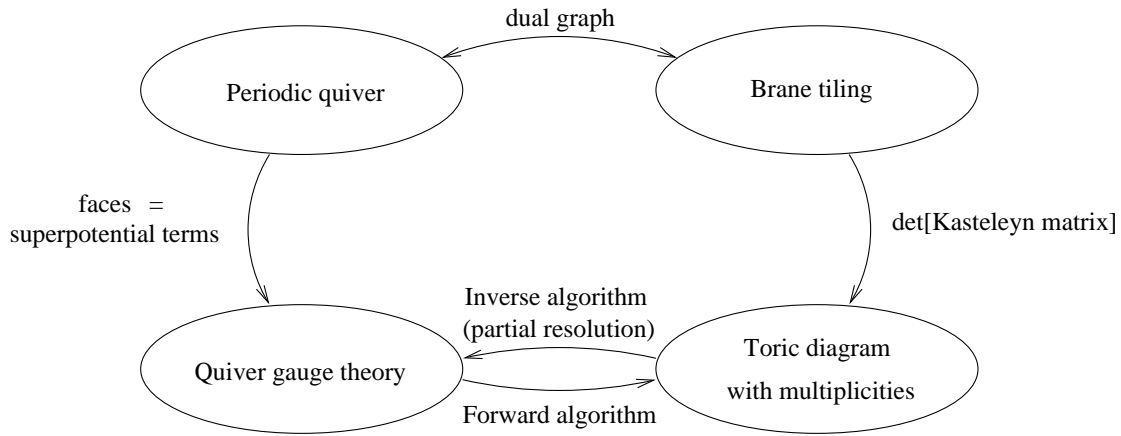}}
  \caption{The logical flowchart.}
  \label{flowchart}
\end{figure}

This is a key observation.   Given the presentation of the quiver data (quiver graph and superpotential) as a planar graph tiling the torus, the bipartite graph appearing in the dimer model (the brane construction of the previous section) is nothing but the planar dual of this graph! Moreover, as we have argued, this dual presentation of the quiver data is {\it physical}, in that it appears directly in string theory as a way to construct the $3+1$-dimensional quiver gauge theory in terms of intersecting NS5 and D5-branes. The logical flow of these ideas is shown in \fref{flowchart}. Some of the concepts in this diagram have not yet been discussed in
this paper, but will be addressed shortly.

Let us see how the properties of the brane tiling arise from those of the quiver theory.
We will show that we can think of the superpotential and quiver together as a tiling
of a two-dimensional surface, where bifundamentals are edges, superpotential
terms are faces, and gauge groups are nodes. We refer to this as the ``periodic quiver'' representation. 
The toric condition, which states that each matter field appears in precisely two superpotential terms of opposite sign, means that the faces all glue together in pairs along the common edges.  Since every field is represented exactly twice in the superpotential, this tiling has no boundaries.  Thus, the quiver and its superpotential may be combined to give a tiling of a Riemann surface without boundary; this periodic quiver gives a discretization of the torus. Since the Euler characteristic of the quiver is zero for toric theories (as discussed in the previous section), the quiver and superpotential data are equivalent to a planar tiling of the two-dimensional torus.   
See Figure 5 of
\cite{Feng:2000mi} for an early example of a periodic quiver.

This tiling has additional structure.  The toric condition implies that adjacent faces of the tiling may be labelled with opposite signs according to the sign of the corresponding term in the superpotential.  Thus, under the planar duality the {\it vertices} of the dual graph may be labelled with opposite signs; this is the bipartite property of the dimer model.   Since the periodic quiver is defined on the torus, the dual bipartite graph also lives on the torus.

Anomaly cancellation of the quiver gauge theory is represented by the balancing of all incoming and outgoing arrows at every node of the quiver.  In the dual graph, bipartiteness means that the edges carry a natural orientation (e.g.~from black to white).  This induces an orientation for the dual edges, which transition between adjacent faces of the brane tiling (vertices of the planar quiver).  For example, these dual arrows point in a direction such that, looking at an arrow from its tail to its head, the black node is to the left and the white node is to the right (this is just a convention and the opposite choice is equivalent by charge conjugation). Arrows around a face in $G$ alternate between incoming and outcoming arrows of the quiver; this is how anomaly cancellation is manifested in the brane tiling picture. Alternatively, we can say that arrows ``circulate'' clockwise around white nodes and counterclockwise around black nodes.

\begin{figure}[ht]
  \epsfxsize = 10cm
  \centerline{\epsfbox{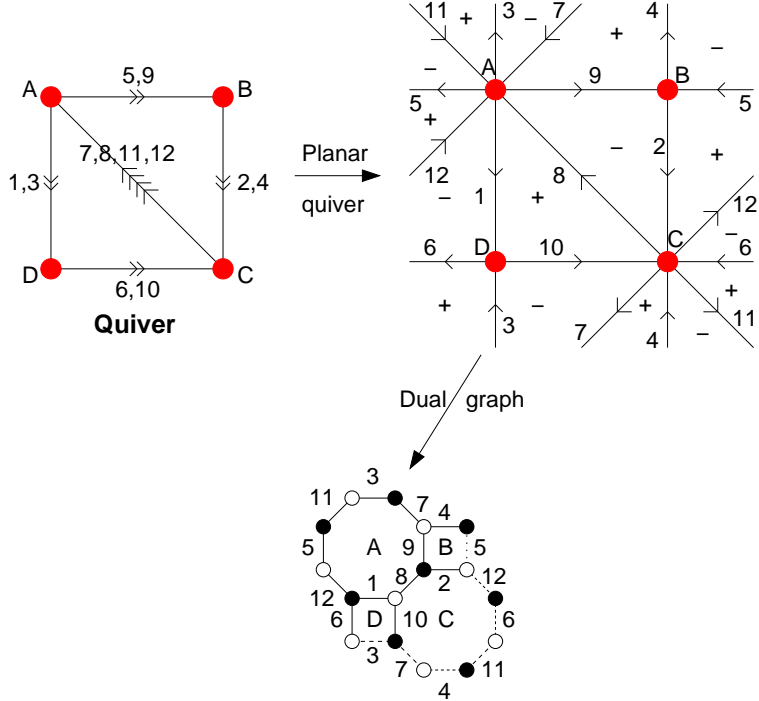}}
  \caption{The quiver gauge theory associated to one of the toric
phases of the cone over ${\bf F}_0$.  In the upper right the quiver and
superpotential \eref{superpot.f0} are combined into the periodic
quiver defined on $T^2$.  The terms in the superpotential bound the
faces of the periodic quiver, and the signs are indicated and have the
dual-bipartite property that all adjacent faces have opposite sign.
To get the bottom picture, we take the planar dual graph and indicate the
bipartite property of this graph by coloring the vertices
alternately.  The dashed lines indicate edges of the graph that are
duplicated by the periodicity of the torus.  This
defines the brane tiling associated to this $\N=1$
gauge theory.}
  \label{periodic.quiver.f0}
\end{figure}

\fref{periodic.quiver.f0} shows an example of the periodic quiver
construction for the quiver gauge theory associated to one of the
toric phases of the Calabi-Yau cone over ${\bf F}_0$.  The superpotential
for this theory is \cite{Feng:2001xr}

\bea
W &=&X_1 X_{10} X_8 - X_3 X_{10} X_7 - X_2 X_8 X_9 - X_1 X_6 X_{12} \\
\nonumber&+& X_3 X_6 X_{11} + X_4 X_7 X_9 + X_2 X_{12} X_5 - X_4 X_{11} X_5.
\label{superpot.f0}
\eea

\section{Dimer model technology}
\label{sec:dimers}

Given a bipartite graph, a problem of interest to physicists and
mathematicians is to count the number of perfect matchings of the
graph.  A {\bf perfect matching} of a bipartite graph is a subset of
edges (``dimers'') such that every vertex in the graph is an endpoint
of precisely one edge in the set.  A {\bf dimer model} is the
statistical mechanics of such a system, i.e.~of random perfect
matchings of the graph with assigned edge weights.  As discussed in
the previous section, we are interested in dimer models associated to
doubly-periodic graphs, i.e.~graphs defined on the torus $T^2$.
We will now review some basic properties of dimers; for additional
review, see \cite{Hanany:2005ve,Kenyon:2002a}.

Many important properties of the dimer model are governed by the {\bf
Kasteleyn matrix} $K(z,w)$, a weighted, signed adjacency matrix of the
graph with (in our conventions) the rows indexed by the white nodes,
and the columns indexed by the black nodes.  It is constructed as
follows:

To each edge in the graph, multiply the edge weight by $\pm 1$ so that
around every face of the graph the product of the edge weights over
edges bounding the face has the following sign

\begin{equation}
\mbox{sign}(\prod_i e_i) = \left\{\begin{array}{cc}
+1 &  {\mbox \it if} \ (\mbox{\# edges}) = 2 \mod 4 \\
-1 & {\mbox \it if} \ (\mbox{\# edges}) = 0 \mod 4
\end{array}
\right.
\label{eq:signs}
\end{equation}
It is always possible to arrange this \cite{Kasteleyn}.

The coloring of vertices in the graph induces an orientation to the
edges, for example the orientation ``black'' to ``white''.  This
orientation corresponds to the orientation of the chiral multiplets of
the quiver theory, as discussed in the previous section.  Now construct
paths $\gamma_w$, $\gamma_z$ in the dual graph (i.e.~the periodic
quiver) that wind once around the $(0,1)$ and $(1,0)$ cycles of the
torus, respectively.  We will refer to these fundamental paths as {\bf flux lines}.
In terms of the periodic quiver, the paths
$\gamma$ pick out a subset of the chiral multiplets whose product is
gauge-invariant and forms a closed path that winds around one of the
fundamental cycles of the torus.  For every such edge (chiral
multiplet) in $G$ crossed by $\gamma$, multiply the edge weight by a
factor of $w$ or $1/w$ (respectively $z$, $1/z$) according to the
relative orientation of the edges in $G$ crossed by $\gamma$.

The adjacency matrix of the graph $G$ weighted by the above factors is
the Kasteleyn matrix $K(z,w)$ of the graph.  The determinant of this
matrix $P(z,w) = \det K$ is a Laurent polynomial (i.e.~negative powers
may appear) called the characteristic polynomial of the dimer model

\beq
P(z,w)=\sum_{i,j} c_{ij} z^iw^j.
\eeq
This polynomial provides the link between dimer models and toric geometry \cite{Hanany:2005ve}.

Given an arbitrary ``reference'' matching $M_0$ on the graph, for any
matching $M$ the difference $M - M_0$ defines a set of closed curves
on the graph in $T^2$. This in turn defines a height function on the
faces of the graph: when a path in the dual graph crosses the curve,
the height is increased or decreased by 1 according to the orientation
of the crossing.  A different choice of reference matching $M_0$
shifts the height function by a constant. Thus, only differences in
height are physically significant.

In terms of the height function, the characteristic polynomial takes
the following form:

\begin{equation}
\label{eq:det}
P(z,w) = z^{h_{x0}} w^{h_{y0}} \sum c_{h_x,h_y} (-1)^{h_x + h_y + h_x h_y} z^{h_x} w^{h_y}
\end{equation}
where $c_{h_x,h_y}$ are integer coefficients that count the number of
paths on the graph with height change $(h_x,h_y)$ around the two
fundamental cycles of the torus.

The overall normalization of $P(z,w)$ is not physically meaningful:
since the graph does not come with a prescribed embedding into the
torus (only a choice of periodicity), the paths $\gamma_{z,w}$ winding
around the primitive cycles of the torus may be taken to cross any
edges en route.  Different choices of paths $\gamma$ multiply the
characteristic polynomial by an overall power $z^i w^j$, and by an
appropriate choice of path $P(z,w)$ can always be normalized to
contain only non-negative powers of $z$ and $w$.

The {\bf Newton polygon} $N(P)$ is a convex polygon in $\Z^2$
generated by the set of integer exponents of the monomials in $P$.
In \cite{Hanany:2005ve}, it was conjectured that the Newton polygon 
can be interpreted as the toric diagram associated to the moduli space of 
the quiver gauge theory, which by assumption is a non-compact toric Calabi-Yau
3-fold. In the following section, we will
prove that the perfect matchings of the dimer model are in 1-1
correspondence with the fields of the gauged linear
sigma model that describes the probed toric geometry.

The connection between dimer models and toric geometry was explored in
\cite{Hanany:2005ve}. In that paper the action of orbifolding the
toric singularity was understood in terms of the dimer model: the
orbifold action by $\Z_m \times \Z_n$ corresponds to enlarging the
fundamental domain of the graph by $m \times n$ copies, and
non-diagonal orbifold actions correspond to a choice of periodicity of
the torus, i.e.~an offset in how the neighboring domains are
adjoined. Furthermore, results analogous to 
the Inverse Algorithm were developed for studying
arbitrary toric singularities and their associated quiver theories.
The present paper
derives and significantly extends the results of
\cite{Hanany:2005ve}, and places them into the context of string
theory.

\begin{figure}[ht]
  \epsfxsize = 15cm
  \centerline{\epsfbox{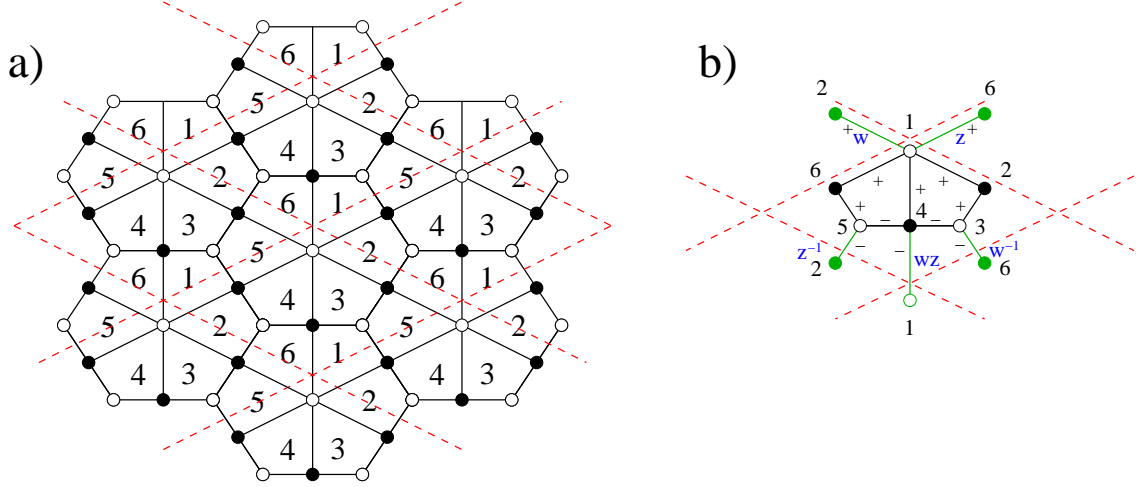}}
  \caption{a) Brane tiling for Model I of ${\bf dP}_3$ with flux lines indicated in red. b) Unit cell  for Model I of ${\bf dP}_3$. We show the edges connecting to images of the fundamental nodes in green. We also indicate the signs associated to each edge as well as the powers of $w$ and $z$ corresponding to crossing flux lines.}
  \label{fluxes_cell_dP3_1}
\end{figure}

Let us illustrate how the computation of the Kasteleyn matrix and the
toric diagram works for the case of Model I of ${\bf dP}_3$. The brane
configuration is shown in \fref{fluxes_cell_dP3_1}a. The corresponding
unit cell is presented in \fref{fluxes_cell_dP3_1}b. As expected, it contains
one valence 6, three valence 4 and two valence 3 nodes. It also
contains twelve edges, corresponding to the twelve bifundamental
fields in the quiver.

\begin{figure}[ht]
  \epsfxsize = 5cm
  \centerline{\epsfbox{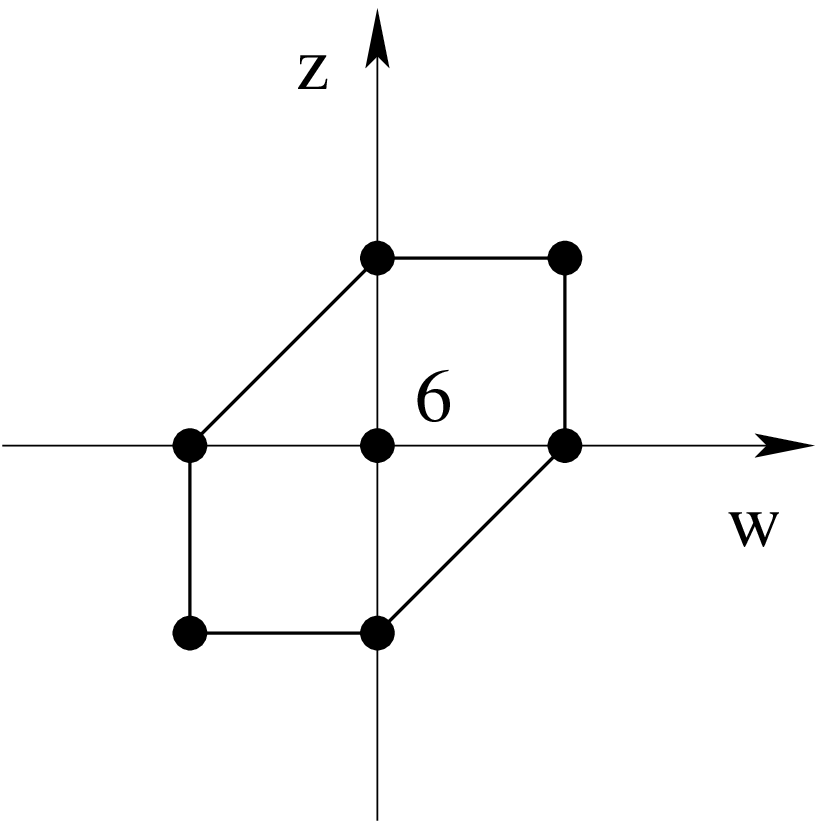}}
  \caption{Toric diagram for Model I of ${\bf dP}_3$ derived from the characteristic polynomial in \eref{pol_dP3_1}.}
  \label{toric_dP3_1}
\end{figure}

From the unit cell, we derive the following Kasteleyn matrix

\beq
K=\left( \begin{array}{c|ccc}  & \ \ 2 \ \ & \ \ 4 \ \ & \ \ 6 \ \ \\ \hline
                             1 & 1+w & 1-zw & 1+z \\
                             3 & 1   & -1 & -w^{-1} \\
                             5 & -z^{-1} & -1 & 1 \end{array} \right)
\label{K_dP3_1}
\eeq
We observe that is has twelve monomials, associated to the twelve bifundamental fields. This matrix leads to the characteristic polynomial
\beq
P(z,w)=w^{-1}z^{-1}-z^{-1}-w^{-1}-6-w-z+wz.
\label{pol_dP3_1}
\eeq
The toric data corresponding to this gauge theory can be read from this polynomial, and is shown in \fref{toric_dP3_1}.

The Kasteleyn matrix is a square matrix whose size is equal to half the total number of points in the unit cell. Thus, for a given toric quiver $K$ is a  $N_W/2 \times N_W/2$ matrix. This is remarkable, since this size can be very modest even for very complicated gauge theories. The simplicity of computing the toric data using this procedure should be contrasted with the difficulty of the Forward Algorithm.

This procedure has a profound impact on the study of quiver theories for arbitrary toric singularities. Given a candidate quiver theory for D3-branes over some geometry, instead of running the lengthy Forward Algorithm, one simply constructs the associated brane tiling using the rules of Section \ref{section_tilings} and computes the corresponding characteristic polynomial. We can thus refer to the determination of toric data from  brane tilings as the {\bf Fast Forward Algorithm}\footnote{A name coined by Pavlos Kazakopoulous.}. This simplification will become clear when we present explicit results for infinite families of arbitrarily large quivers in Sections \ref{sec:infquivers}.

\section{An explicit correspondence between dimers and GLSMs}

\label{section_dimers_GLSM}

Following \cite{Hanany:2005ve}, we have argued in the previous section that the characteristic polynomial encodes the toric data of the probed geometry.
We now explore the reason for this connection, establishing a correspondence between fields in the gauged linear sigma model
description of the singularity and perfect matchings in the brane tiling.

Given a toric Calabi-Yau 3-fold, the principles of determining the gauge theory on the world-volume of a stack of D3-brane probes are well established. 
Conversely, the determination of the toric data of the singularity from the gauge theory is also clear. This procedure has been algorithmized in 
\cite{Feng:2000mi} and dubbed the {\bf Forward Algorithm}. Nevertheless, although a general prescription exists, its applicability beyond the simplest 
cases is limited due to the computational complexity of the algorithm. 

Let us review the main ideas underlying the Forward Algorithm (for a detailed description and explicit examples, we refer the reader to 
\cite{Feng:2000mi}). The starting point is a quiver with $r$ $SU(N)$ gauge groups and  bifundamentals $X_i$, $i=1,\ldots,m$, together with a 
superpotential. The toric data that describes the probed geometry is computed using the following steps:

\begin{itemize}
\item Use F-term equations to express all bifundamental fields $X_i$ in terms of $r+2$ independent variables $v_j$. 
The $v_j$'s can be simply equal to a subset of the bifundamentals.
The connection between these variables and the original bifundamental fields is encoded in an 
$m \times (r+2)$ matrix $K$ (this matrix should not be confused with the Kasteleyn matrix;
which of them we are talking about will be clear from the context), such that 

\beq
X_i=\prod_j v_j^{K_{ij}}, \ \ \ \ i=1,2,\ldots, m, \ \ \ \ j=1,2,\ldots, r+2.
\eeq

Since the F-term equations take the form of a monomial equated to another monomial, it is clear that generically $K_{ij}$ has negative 
entries (i.e.~negative powers of the $v_j$ can appear in the expressions for the $X_i$).

\item In order to avoid the use of negative powers, a new set of variables $p_\alpha$, $\alpha=1,\ldots,c$, is introduced. 
The number $c$ is not known {\it a priori} in this approach, and must be determined as part of the algorithm. We will later 
see that it corresponds to the number of perfect matchings of $G$, the periodic bipartite graph dual to the quiver.

\item The reduction of the $c$ $p_\alpha$'s to the $r+2$ independent variables $v_i$ is achieved by introducing a $U(1)^{c-(r+2)}$ 
gauge group. The action of this group is encoded in a $(c-r-2) \times c$ charge matrix $Q$.

\item The original $U(1)^{r-1}$ action (one of the $r$ $U(1)$'s is redundant) determining the D-terms is recast in terms
of the $p_\alpha$ by means of a $(r-1) \times c$ charge matrix $Q_D$.
 
\item $Q$ and $Q_D$ are combined in the total matrix of charges $Q_t$. The $U(1)$ actions of the symplectic quotient defining the 
toric variety correspond to a basis of linear relations among the vectors in the toric diagram. Thus, the toric diagram 
corresponds to the columns in a matrix $G_t$ such that $G_t=(\mbox{ker} \ Q_t)^T$.

\end{itemize}

At this stage, it is important to stress some points. The main difficulty in the Forward Algorithm is the computation of $T$, which is used 
to map the intermediate variables $v_i$ to the GLSM fields $p_\alpha$. Its determination involves the computation of a dual cone, consisting 
of vectors such that $\vec{K} \cdot \vec{T} \geq 0$. The number of operations involved grows drastically with the ``size'' (i.e.~the number of nodes 
and bifundamental fields) of the quiver. The computation becomes prohibitive even for quivers of moderate complexity. Thus, one is forced to 
appeal to alternative approaches such as (un-)Higgsing \cite{Feng:2002fv}. Perhaps the most dramatic examples of this limitation are provided by 
recently discovered infinite families of gauge theories for the $Y^{p,q}$ \cite{Benvenuti:2004dy} and $X^{p,q}$ \cite{Hanany:2005hq} singularities. 
The methods presented in this section will enable us to treat such geometries.  This also represents a significant improvement over the brute force 
methods of \cite{Hanany:2005ve}, since the relevant brane tiling may essentially be written down directly from the data of the quiver theory.

It is natural to ask whether the possibility of associating dimer configurations to a gauge theory, made possible due to the introduction of brane 
tilings, can be exploited to find a natural set of variables playing the role of the $p_\alpha$'s, overcoming the main intricacies of 
the Forward Algorithm. This is indeed the case, and we now elaborate on the details of the {\bf dimer/GLSM correspondence}. The fact
that the GLSM multiplicities are counted by the $c_{ij}$ coefficients in the characteristic polynomial provides some motivation
for the correspondence.

We denote the perfect matchings as $\tilde{p}_\alpha$. Every perfect matching corresponds to a 
collection of edges in the tiling. Hence, we can define a natural product between an edge $e_i$, corresponding to a bifundamental
field $X_i$, and a perfect matching $\tilde{p}_\alpha$

\beq
<e_i,\tilde{p}_\alpha>=\left\{ \begin{array}{l} 1 \mbox{ if } e_i \in \tilde{p}_\alpha \\
                                                      0 \mbox{ if } e_i \notin \tilde{p}_\alpha  \end{array} \right.
\label{projection_dimers}
\eeq

Given this product, we propose the following mapping between bifundamental fields and the perfect matching variables $\tilde{p}_\alpha$

\beq
X_i=\prod_\alpha \tilde{p}_\alpha^{<e_i,\tilde{p}_\alpha>}.
\label{X_perfect_matching}
\eeq

\begin{figure}[ht]
  \epsfxsize = 12.5cm
  \centerline{\epsfbox{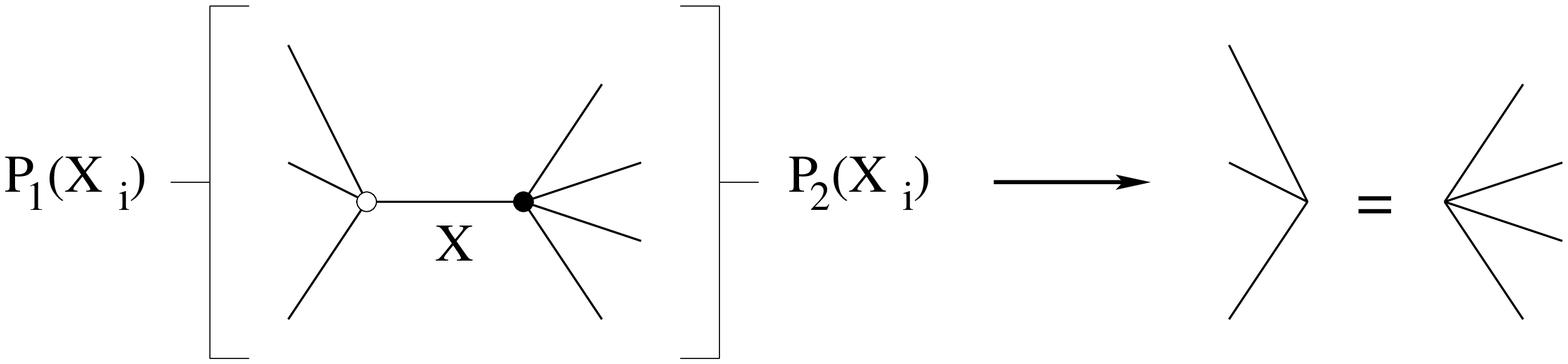}}
  \caption{F-term equations from the brane tiling perspective.}
  \label{F-term}
\end{figure}

According to \eref{projection_dimers}, the $X_i$ involve only possitive powers of the $\tilde{p}_\alpha$.
We will now show that F-term equations are trivially satisfied when the bifundamental fields are expressed in terms of 
perfect matchings variables according to \eref{X_perfect_matching}. For any given bifundamental field $X_0$, we have

\beq
W=X_0 P_1(X_i)-X_0 P_2(X_i)+\ldots
\eeq
where we have singled out the two terms in the superpotential that involve $X_0$. $P_1(X_i)$ and
$P_2(X_i)$ represent products of bifundamental fields. The F-term equation
associated to $X_0$ becomes

\beq
\partial_{X_0} W = 0 \ \ \ \ \Leftrightarrow \ \ \ \ P_1(X_i)=P_2(X_i).
\label{F_term_1}
\eeq
This condition has a simple interpretation in terms of the bipartite graph, as shown in \fref{F-term}.

After excluding the edge associated to $X_0$, the product of edges connected to node 1 has to be 
equal to the product of edges connected to node 2. In terms of perfect matchings, \eref{F_term_1}
becomes

\beq
\prod_{i \in P_1} \prod_\alpha \tilde{p}_\alpha^{<e_i,\tilde{p}_\alpha>}=
\prod_{i \in P_2} \prod_\alpha \tilde{p}_\alpha^{<e_i,\tilde{p}_\alpha>}.
\label{F_term_2}
\eeq
Every time that a given $\tilde{p}_\alpha$ appears on the L.H.S. of \eref{F_term_2}, it has to
appear on the R.H.S. Here is where the fact that the $\tilde{p}_\alpha$'s are perfect matchings
becomes important: since nodes 1 and 2 are separated exactly by one edge (the one corresponding
to $X_0$) every time a perfect matching contains any of the edges in $P_1$, it contains one of the edges
in $P_2$. This is necessary for the $p_\alpha$ to be a perfect matching (nodes 1 and 2 have to be covered
exactly once). Thus, perfect matchings are the appropriate choice of variables that satisfy
F-term conditions automatically. We conclude that the perfect matchings can be identified with the GLSM 
fields $\tilde{p}_\alpha=p_\alpha$. Then, the matrix that maps the bifundamental fields to the 
GLSM fields is

\beq
(KT)_{i\alpha}=<e_i,\tilde{p}_\alpha>.
\label{KT}
\eeq

\subsection{A detailed example: the Suspended Pinch Point}

\label{sec:spp}

\begin{figure}[ht]
  \epsfxsize = 3cm
  \centerline{\epsfbox{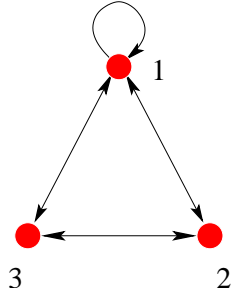}}
  \caption{Quiver diagram for the SPP.}
  \label{quiver_SPP}
\end{figure}

Let us illustrate the simplifications achieved by identifying GLSM fields with perfect matchings with an explicit example. To do
so, we choose the Suspended Pinch Point (SPP) \cite{Morrison:1998cs}. The SPP has a quiver shown in \fref{quiver_SPP} 
with superpotential 

\beq
W=X_{21}X_{12}X_{23}X_{32} - X_{32}X_{23}X_{31}X_{13} + X_{13}X_{31}X_{11} - X_{12}X_{21}X_{11}.
\eeq

\begin{figure}[ht]
  \epsfxsize = 5.5cm
  \centerline{\epsfbox{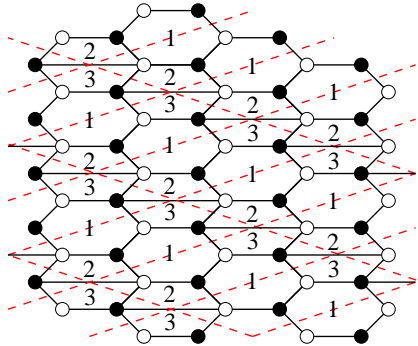}}
  \caption{Brane tiling for the SPP.}
  \label{dimer_SPP}
\end{figure}

\begin{figure}[ht]
  \epsfxsize = 3.5cm
  \centerline{\epsfbox{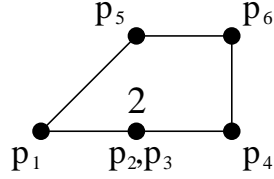}}
  \caption{Toric diagram for the SPP. We indicate the perfect matchings corresponding to each node in the toric diagram.}
  \label{toric_SPP}
\end{figure}

It is interesting to see how our methods apply to this example, since it has both adjoint fields and bidirectional arrows. 
\fref{dimer_SPP} shows the brane tiling for the SPP. The adjoint field in the quiver corresponds to an edge between two faces 
in the tiling representing the first gauge group.
The Kasteleyn matrix is 

\beq
K=\left( \begin{array}{c|cc}  & \ \ 2 \ \ & \ \ 4 \ \ \\ \hline
                             1 & 1+w^{-1} & z+w^{-1}z \\

                             3 & 1 & 1+w^{-1}  \end{array} \right)
\eeq
from which we determine the characteristic polynomial

\beq
P(z,w)=w^{-2}+2w^{-1}+1-w^{-1}z-z.
\eeq
From it, we construct the toric diagram shown in \fref{toric_SPP}. 

\begin{figure}[ht]
  \epsfxsize = 14cm
  \centerline{\epsfbox{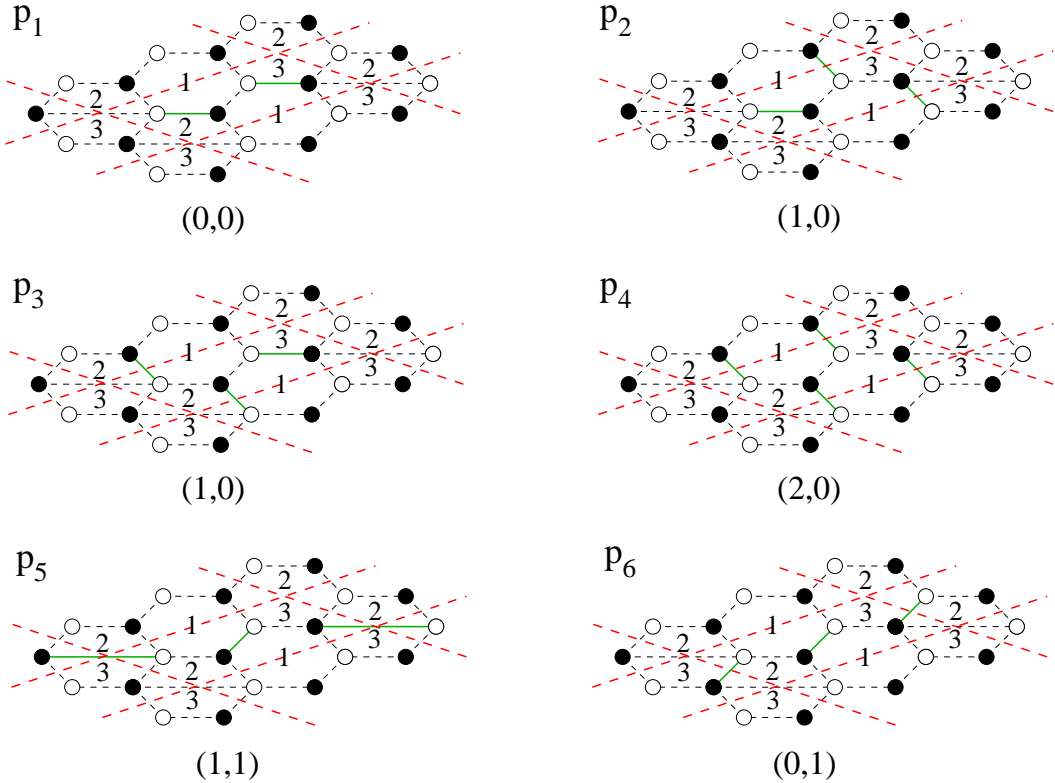}}
  \caption{Perfect matchings for the SPP. We indicate the slopes $(h_w,h_z)$, which allow the identification of the corresponding
node in the toric diagram as shown in \fref{toric_SPP}.}
  \label{matchings_SPP}
\end{figure}

There are six perfect matchings of the SPP tiling. We show them in \fref{matchings_SPP}. Setting a reference perfect matching, we can 
compute the slope $(h_w,h_z)$ for each of them, i.e.~the height change when moving around the two fundamental cycles of the torus.

Using \eref{projection_dimers} and \eref{KT}, it is straightforward to determine the $KT$ matrix.

\beq
KT=\left( \begin{array}{c|cccccc}  & \ \ p_1 \ \ & \ \ p_2 \ \ & \ \ p_3 \ \ & \ \ p_4 \ \ & \ \ p_5 \ \ & \ \ p_6 \ \ \\ \hline
                           X_{11} & 0 & 0 & 0 & 0 & 1 & 1 \\                             
                           X_{12} & 1 & 1 & 0 & 0 & 0 & 0 \\
                           X_{21} & 0 & 0 & 1 & 1 & 0 & 0 \\
                           X_{31} & 1 & 0 & 1 & 0 & 0 & 0 \\
                           X_{13} & 0 & 1 & 0 & 1 & 0 & 0 \\
                           X_{23} & 0 & 0 & 0 & 0 & 1 & 0 \\
                           X_{32} & 0 & 0 & 0 & 0 & 0 & 1
                               \end{array} \right)
\eeq

This agrees with the computation of this matrix done in Section 3.2 of \cite{Feng:2000mi}.


\section{Massive fields}

\label{section_massive_nodes}

By definition, massive fields appear in the superpotential as quadratic terms.  Therefore they appear in the brane tiling as bivalent vertices.  In the IR limit of the gauge theory, these massive fields become non-dynamical and should be integrated out using their equations of motion.  We now show how to perform this procedure on the brane tiling and Kasteleyn matrix.

By performing a suitable relabelling of fields, one can always write the superpotential as follows (up to an overall minus sign if the quadratic term comes with opposite sign):

\begin{equation}
W(\Phi_i) = \Phi_1 \Phi_2 - \Phi_1 P_1(\Phi_i) - \Phi_2 P_2(\Phi_i) + \ldots
\end{equation}
where the omitted terms do not involve $\Phi_1$, $\Phi_2$, and $P_1$, $P_2$ are products of two disjoint subsets of the remaining $\Phi$ that do not include $\Phi_1$ or $\Phi_2$.  This structure follows from the toric condition, which specifies that each field appears exactly twice in the superpotential, with terms of opposite signs.

\begin{figure}[ht]
  \epsfxsize = 8.5cm
  \centerline{\epsfbox{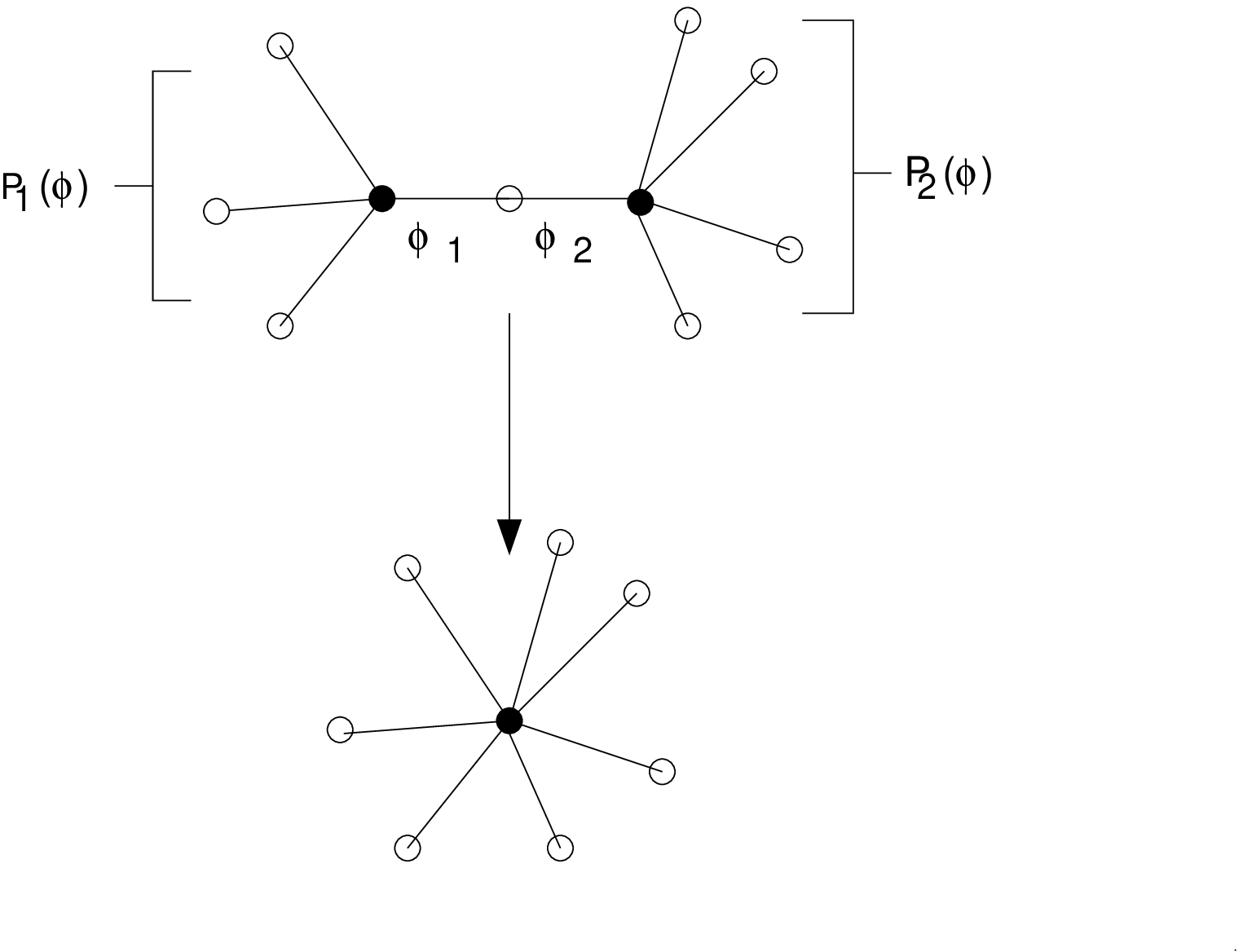}}
  \caption{Integrating out a massive field corresponds to collapsing the two vertices adjacent to a bivalent vertex into a single vertex of higher valence.}
  \label{fig:intout}
\end{figure}

Integrating out $\Phi_1$ and $\Phi_2$ by their equations of motion gives
\begin{equation}
W(\Phi) = - P_1(\Phi_i) P_2(\Phi_i) + \ldots
\end{equation}
This operation takes the form shown in \fref{fig:intout} and collapses two nodes separated by a bivalent node of the opposite color 
into a single node of valence equal to the sum of the valences of the original nodes. 

The operation of integrating out a massive field can also be implemented in terms of the Kasteleyn matrix. From this perspective, it 
is simply row or column reduction of the matrix on rows or columns with two non-zero entries (or a single entry containing two summands, if 
both neighboring vertices to the bivalent vertex are identified in the graph). In the example of figure \ref{fig:intout}, if the bivalent white node has label $1$ 
and the adjacent black nodes are $1'$ and $2'$ (this can always be arranged by a reordering of rows or columns, with the corresponding 
action of $(-1)$ to preserve the determinant), the Kasteleyn matrix (or its transpose) has the following structure:

\begin{equation}
K = \left( \begin{array}{ccccc}
v^{(1)}_{1} & v^{(2)}_{1} &0 &\ldots&0\\
v^{(1)}_{2} & v^{(2)}_{2} & & &\\
\vdots & \vdots & & \star &\\
v^{(1)}_{n} & v^{(2)}_{n} & & &\\
\end{array}\right)
\end{equation}
where $v^{(1)}$ and $v^{(2)}$ index the adjacent nodes to $1'$ and $2'$, i.e.~contain $\deg(P_{1,2}(\Phi)) + 1$ non-zero entries.

Performing elementary column operations\footnote{It is possible that some row and column operations produce a bipartite graph 
corresponding to a gauge theory with different matter content and interactions, but the same IR moduli space. It would be interesting 
to study the physical meaning of these operations in more detail.}, the matrix can be brought to the following form\footnote{If the sets of vertices $v^{(1)}$, $v^{(2)}$ adjacent to vertices $1'$ and $2'$ (excluding the common neighbor $1$) are not disjoint, then after integrating out there will be two or more edges between the same pairs of vertices. In such cases, these multiple edges may be replaced by a single edge carrying the sum of the weights of the individual edges, since this reproduces the correct counting of matchings of the graph.  This is indeed what happens in the column reduction process, which may produce entries that are the sum or difference of two non-zero entries.}:

\begin{equation}
K = \left( \begin{array}{ccccc}
1 & 0 &0 &\ldots&0\\
v^{(1)}_{2}/v^{(1)}_{1} & v^{(2)}_{2} v^{(1)}_{1} -v^{(1)}_{2} v^{(2)}_{1}   & & &\\
\vdots & \vdots & & \star &\\
v^{(1)}_{n}/v^{(1)}_{1} & v^{(2)}_{n} v^{(1)}_{1} -v^{(1)}_{n} v^{(2)}_{1}   & & &\\
\end{array}\right)
\end{equation}
and therefore can be reduced in rank without changing the determinant, by deleting the first row and column, giving the reduced Kasteleyn matrix
\begin{equation}
K = \left( \begin{array}{cccc}
v^{(2)}_{2} v^{(1)}_{1} -v^{(1)}_{2} v^{(2)}_{1}   &\star & \star &  \ldots\\
\vdots & \star & \star & \ldots\\
v^{(2)}_{n} v^{(1)}_{1} -v^{(1)}_{n} v^{(2)}_{1}   &\star &\star &  \ldots\\
\end{array}\right)
\end{equation}
corresponding to the graph with bivalent vertex deleted.

\section{Seiberg duality}
\label{sec:seiberg}

\subsection{Seiberg duality as a transformation of the quiver}

We now discuss how one can understand Seiberg duality from the perspective of the brane tilings. To motivate our construction, let us first recall what happens to a quiver theory when performing Seiberg duality at a single node. This was first done for orbifold quivers in 
 \cite{Schmaltz:1998bg}. Recall first that since Seiberg duality takes a given gauge group $SU(N_c)$ with $N_f$ fundamentals and $N_f$ anti-fundamentals to $SU(N_f - N_c)$, if we want the dual quiver to remain in a toric phase, we are only allowed to dualize on nodes with $N_f = 2N_c$. Dualizing on such a node (call it $I$) is straightforward, and is done as follows:

\begin{itemize}

\item To decouple the dynamics of node $I$ from the rest of the theory, the gauge couplings of the other gauge groups and superpotential should be scaled to zero.  The fields corresponding to edges in the quiver that are not adjacent to $I$ decouple, and the edges between $I$ and other nodes reduce from bifundamental matter to fundamental matter transforming under a global flavor symmetry group.  This reduces the theory to the SQCD-like theory with $2N_c$ flavors and additional gauge singlets, to which Seiberg duality may be applied.

\item Next, reverse the direction of all arrows entering or exiting the dualized node. This is because Seiberg duality requires that the dual quarks transform in the conjugate flavor representations to the originals, and the other end of each bifundamental transforms under a gauge group which acts as an effective flavor symmetry group. Because we want to describe our theory with a quiver, we perform charge conjugation on the dualized node to get back bifundamentals. This is exactly the same as reversing the arrows in the quiver.

\item Next, draw in $N_f$ bifundamentals which correspond to composite (mesonic) operators that are singlets at the dualized node $I$ and carry flavor indices in the pairs  nodes connected to $I$. This is just the usual $Q_i\widetilde Q^j \rightarrow M_i^j$ ``electric quark $\rightarrow$ meson'' map of Seiberg duality, but since each flavor group becomes gauged in the full quiver theory, the Seiberg mesons are promoted to fields in the bifundamental representation of the gauge groups. 

\item In the superpotential, replace any composite singlet operators with the new mesons, and write down new terms corresponding to any new triangles formed by the operators above. It is possible that this will make some fields massive (e.g.~if a cubic term becomes quadratic), in which case the appropriate fields should then be integrated out.
\end{itemize}

\subsection{Seiberg duality as a transformation of the brane tiling}

By writing the action of Seiberg duality in the periodic quiver picture, one may derive the corresponding transformation on the dual brane tiling.  This operation may be encoded in a transformation on the Kasteleyn matrix of the graph, and the recursive application of Seiberg duality may be implemented by computer to traverse the Seiberg duality tree \cite{Franco:2003ja,Franco:2003ea} and enumerate all toric phases
\footnote{Assuming this graph is connected. In fact, this is not allways the case and it is possible for the toric phases to appear in 
disconnected (i.e. connected by non-toric phases) regions of the duality tree. A simple example of this situation is given by the
duality tree of ${\bf dP}_1$. This tree is presented in \cite{Franco:2003ja}, where the connected toric components where denoted ``toric islands''.
In addition, it is interesting to see that if the theory is taken
out of the conformal point by the addition of fractional branes, the cascading RG flow can actually ``migrate" 
among these islands \cite{Franco:2004jz}.}.

Consider a node in the periodic quiver.  For the toric phases of the quiver all nodes in the quiver correspond to gauge groups of equal rank.  If the node has 2 incoming arrows (and therefore 2 outgoing arrows by anomaly cancellation, for a total of 4 arrows), then for this gauge group $N_f = 2 N_c$, and Seiberg duality maps

\begin{equation}
N_c \mapsto \hat{N_c} = N_f - N_c = N_c
\end{equation}
so after the duality the theory remains in a toric phase.

\begin{figure}[ht]
  \epsfxsize = 10cm
  \centerline{\epsfbox{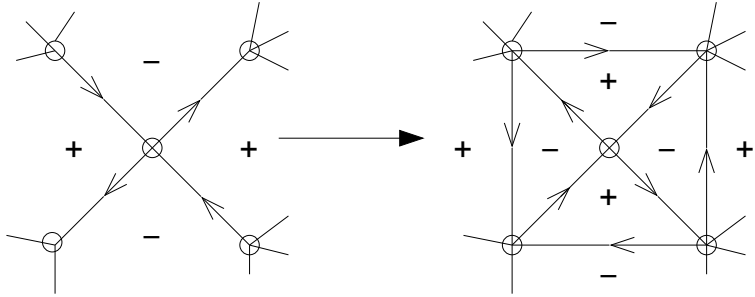}}
  \caption{The action of Seiberg duality on a periodic quiver to produce another toric phase of the quiver.  Also marked are the signs of superpotential terms, showing that the new terms (faces) are consistent with the pre-existing 2-coloring of the global graph.}
  \label{seiberg.quiver}
\end{figure}

At such a node $V$, a generic quiver can be represented as in \fref{seiberg.quiver}.  The 4 faces $F_i$ adjacent to $V$ share an edge with their adjacent faces, and contain some number of additional edges.  

The neighboring vertices to $V$ are not necessarily all distinct (they may be identified by the periodicity of the torus on which the quiver lives).  However by the periodic quiver construction, if there are multiple fields in the quiver connecting the same two vertices, these appear as distinct edges in the periodic quiver.

Note that the new mesons can only appear between adjacent vertices in the planar quiver, because the edges connecting opposing vertices do not have a compatible orientation, so they cannot form a holomorphic, gauge-invariant combination.  There are indeed $4$ such arrows that can be drawn on the quiver corresponding to the $2\times 2 = 4$ Seiberg mesons.

\begin{figure}[ht]
  \epsfxsize = 10cm
  \centerline{\epsfbox{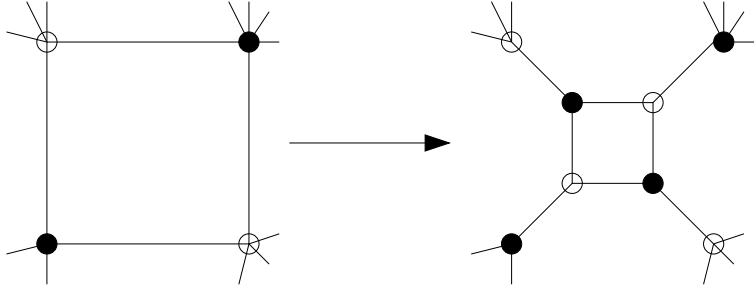}}
  \caption{Seiberg duality acting on a brane tiling to produce another toric phase.  This is the planar dual to the operation depicted in 
\fref{seiberg.quiver}. Whenever 2-valent nodes are generated by this transformation, the corresponding massive fields can be integrated out
as explained in Section \ref{section_massive_nodes}.}
  \label{seiberg.dimer}
\end{figure}

It is easy to translate this operation to the dual brane tiling.  Gauge groups with $N_f = 2 N_c$ correspond to quadrilaterals in the tiling.  Performing Seiberg duality on such a face corresponds to the operation depicted in \fref{seiberg.dimer}\footnote{The extension to non-toric Seiberg dualities appears obvious on the periodic quiver, although there are subtleties involved in the precise operation on the Kasteleyn matrix.}.

Note that this operation (and the dual operation on the quiver) are local operations on the graph, in that they only affect a face and its neighbors, and the global structure of the graph is unaffected\footnote{The transformation that we have identified with the action of Seiberg duality on the bipartite graph was discussed in \cite{Propp:2001a}, where it was referred to as ``urban renewal''.  This work used a different assignment of weights in the transformed graph in order to keep the determinant (i.e.~the GLSM field multiplicities, in our language) invariant across the operation.  This is not what we want for Seiberg duality, which maps a toric diagram with one set of multiplicities to the same toric diagram with (in general) {\it different} multiplicities.}.

\begin{figure}[ht]
\epsfxsize = 10cm
\centerline{\epsfbox{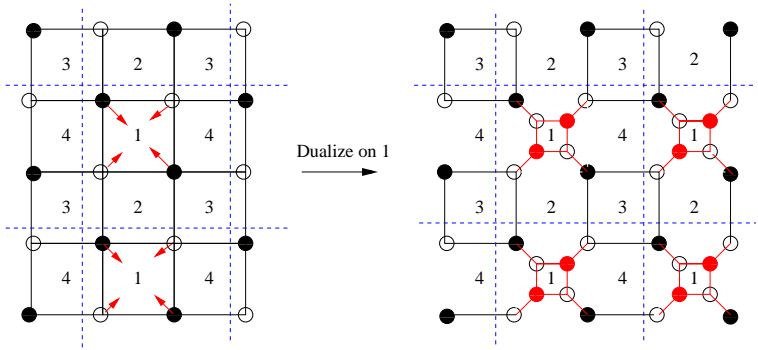}}
\caption{The operation of Seiberg duality on a phase of ${\bf F}_0$.}
\label{f0dual}
\end{figure}

As a simple example, consider ${\bf F}_0$. This is a $\IZ_2$ orbifold
of the conifold, and as such one can simply take the two-cell
fundamental domain of the conifold and double its area (with an 
appropriate choice of periodicity) to get 
the ${\bf F}_0$ fundamental domain; this phase of this theory is
given by a square graph with four different cells. In \fref{f0dual},
we have drawn this phase of the theory as well as the phase obtained by
dualizing on face 1. The blue dotted lines are the lines of magnetic
flux delineating the fundamental region, which do not change during
Seiberg duality.  It is straightforward to see that these regions give
the correct Kasteleyn matrices, and reproduce the known multiplicities
of sigma model fields \cite{Hanany:2005ve}.

It is useful to see how this action of Seiberg duality can be understood from the brane perspective. Since the area of each cell (volume of the D-brane) is related to the gauge coupling of the corresponding group \cite{Hanany:1998ru}, one would expect that Seiberg duality could be viewed as a cell shrinking and then growing with the opposite orientation, e.g.~as branes move through one another. It is possible to see this from \fref{f0dual}: we can simply take the NS5-branes at the sides of region 1 and pull them through one another. In doing this, we generate the diagonal lines.  Since we are in a toric phase with $N_f = 2N_c$, the ranks of the gauge groups do not change in this crossing operation and no new branes are created.

\subsection{Seiberg duality acting on the Kasteleyn matrix}

Since the Kasteleyn matrix encodes all of the information about the graph, it is possible to implement the transformation of Seiberg duality directly in terms of the matrix.  The first step is to identify the candidate (quadrilateral) faces to be dualized.  These form a square in the Kasteleyn matrix, e.g.~:

\begin{equation}
K = \left(\begin{array}{ccccc}
\star & a & \star &\ldots & b \\
\star & \star & \star &\ldots & \star \\
\vdots & \vdots & \vdots &&\vdots\\
\star & c & \star &\ldots & d \\
\star & \star & \star &\ldots & \star \\\end{array}\right)
\end{equation}
However not all such squares represent the boundary of a face, e.g.~on small enough graphs there can be a closed path of 4 edges which winds around the torus (for a closed cycle of 4 edges there are no other possibilities, as there is no room for additional ``internal" edges that would allow the cycle to have zero winding but not bound a face of the graph).  The way to distinguish these cases is to use the magnetic flux through the cycle; if the cycle has no net winding around the torus then the flux lines $\gamma_z, \gamma_w$ must each cross the cycle twice: once to enter and once to leave.  Depending on the orientation with which they cross the edges (which depends on the choice of paths $\gamma$ and is therefore not invariant), they may each contribute $z$ or $1/z$ (similarly $w$ or $1/w$), but it is invariantly true that the product of the edges must have even degree in both $z$ and $w$.  Conversely, a path with net winding around the torus will have odd degree in one or both of $z$ and $w$.

Having identified the four edges forming the quadrilateral to be dualized, we wish to implement the transformation on the underlying graph depicted in \fref{seiberg.dimer}.  This requires the addition of 2 white and 2 black nodes to the graph, increasing the rank of the adjacency matrix by 2.   The large square is removed from the graph by setting to zero the four edges $a, b, c, d$ found previously, the smaller square is added in by setting to non-zero the weights in the new $2 \times 2$ diagonal block, and the new smaller square is connected to the rest of the graph by adding non-zero elements to the $2 \times n$ and $n \times 2$ blocks in the rows and columns corresponding to the removed entries.

Since the new square has opposite orientation with respect to the large square, the power of  $z$ and $w$ along the edges must be inverted.  This may change the normalization of the determinant, so to correct this we can rescale a row by $z^{d_z}$ and a column by $w^{d_w}$, where $d_{w,z} = \deg_{w,z} a b c d$.  Finally, since the graph transformation adds 2 additional edges to each of the 4 faces adjacent to the square, each of these faces must gain an additional minus sign on one of the new edges bounding it, in order to satisfy the sign rules discussed in section \ref{sec:dimers}.

Explicitly, the new Kasteleyn matrix corresponding to the Seiberg dual graph can be written in the following form (where for convenience we have relabeled the rows and columns to bring the 4 entries corresponding to edges of the quadrilateral into the bottom-right position):

\beq
K=\left( \begin{array}{|c|ccc|} \hline 
& & & \\
A_{(n-2) \times (n-2)} & & B_{(n-2) \times 2} & \\ 
& & &\\ \hline
C_{2 \times (n-2)} &        & \begin{array}{cc}a & b\\
c & d\end{array}&\\
\hline \end{array}\right) \ \ \ \mapsto \ \ \ 
K'=\left( \begin{array}{|c|c|c|} \hline
& &\\
A_{(n-2) \times (n-2)} & B_{(n-2) \times 2} &0_{2 \times 2} \\ 
& &\\ \hline
C_{2 \times (n-2)} \ \ & 0_{2 \times 2} & \begin{array}{cc}
z^{d_z} &0 \\ 
0 & 1\\
\end{array}\\
\hline
0_{2 \times (n-2)} &\begin{array}{cc}0 & -1\\ -w^{d_w} & 0 \\\end{array}& \begin{array}{cc}\frac{1}{b} z^{d_z} & \frac{1}{d}\\\frac{1}{a} w^{d_w} z^{d_z} & \frac{1}{c} w^{d_w}\end{array}\\
\hline
\end{array}\right)
\eeq
As discussed above, after Seiberg duality there may be massive fields in the theory which should be integrated out; we described in section \ref{section_massive_nodes} how to implement this on the Kasteleyn matrix.

This form of Seiberg duality is amenable to efficient implementation on computer.  
For example, the enumeration of 17 of the toric phases of the $Y^{6,0}$ quiver took several days using the algorithm of \cite{Hanany:2005ve} (which was itself much more efficient than the previous Inverse Algorithm).  Since $Y^{6,0}$ is a $\IZ_6$ orbifold of the conifold, it is possible to use the orbifold formul\ae\ in \cite{Hanany:2005ve} and immediately write down the Kasteleyn matrix for two of these phases; either one is a suitable starting point for iteration of Seiberg duality, which produces these 17 phases within a few seconds.

However, this raises an important subtlety: it is possible for two distinct quiver theories to have the same set of multiplicities of points in the toric diagram.  It had previously been conjectured that the multiplicities of GLSM fields uniquely characterized the possible toric phases of the quiver gauge theory, i.e.~in dimer language that the characteristic polynomial was an invariant of the dimer graph up to graph isomorphism (relabelling of fields).  In the example of $Y^{6,0}$, only 17 distinct sets of multiplicities are produced, compared with an expectation of 18 toric phases \cite{Hanany:2005hq}; this mismatch was noted already in \cite{Hanany:2005ve}\footnote{A simpler example manifesting this collision of multiplicities is that of pseudo-del Pezzo 5: two of the four toric phases of this theory have the same multiplicities, as we discuss in section \ref{sec:examples}.}.  Furthermore, extending the set of data considered to include the set of orders of terms in the superpotential (which can be read off from the Kasteleyn matrix independently of the field labelling), and the set of neighboring toric phases that are reachable under Seiberg duality, still only distinguishes 17 distinct phases.  By writing down the brane tilings explicitly and reconstructing the quivers\footnote{This work was done in conjunction with P. Kazakopoulos.}, we were able to isolate the ``missing'' 18th phase and confirm that it indeed has the same toric diagram with multiplicities as one of the remaining 17, but is nonetheless a distinct quiver theory that is not equivalent under field redefinition. In addition, these  
two phases have the same superpotential orders and Seiberg duality neighbors. In Section \ref{section_PdP_5}, we will present a similar example in which 
two gauge theories produce the same toric diagram and multiplicities, although in that case the superpotentials will have different orders
and numbers of terms.

How can we understand this situation?  It is a known result in mathematics that the characteristic polynomial (in the usual linear algebra sense) of the adjacency matrix of a graph does not uniquely characterize the graph up to isomorphism: there may be two distinct graphs with the same characteristic polynomial.  The determinant of the Kasteleyn matrix of the graph is essentially a double characteristic polynomial (due to the block structure of the matrix, as explained in \cite{Hanany:2005ve}), so the result explains the observed non-uniqueness of GLSM multiplicities of the toric quivers.  Similarly, it is believed that there exists {\it no} invariant of a graph up to isomorphism that distinguishes between all non-isomorphic graphs.  In other words, the only 
invariant of a graph that characterizes it uniquely is the graph itself, and in order to distinguish between the pathological cases where the would-be-invariants fail, one must resort to explicit testing of graph isomorphism, which is an expensive (non-polynomial) operation.  Thus, the enumeration of toric phases by testing graph invariants such as the characteristic polynomial can only produce a lower bound on the number of phases, and in general there may be phases (or even entire regions of the toric duality graph) that are missed by this counting.

\section{Partial resolution}

\label{section_partial_resolution}

Many of the first known examples of gauge theories dual to toric geometries
were described by embedding them in orbifolds \cite{Feng:2000mi,Feng:2001xr,Beasley:2001zp}.  For example, partial resolutions of  $\IC^3/\IZ_3 \times \IZ_3$ give the first three del Pezzo theories and ${\bf F}_0$, among others. Partially resolving the orbifold singularity corresponds to turning on Fayet-Iliopoulos terms in the dual gauge theory, which by the D-flatness conditions gives vacuum expectation values to bifundamental fields. These vevs then reduce the rank of the gauge group via the Higgs mechanism.  From the standpoint of the toric diagram, this is simply removing an external point. Doing so decreases the area of the toric diagram, and consequently decreases the number of gauge groups in the dual superconformal theory. 

It is straightforward to see how Higgsing operates from the
perspective of the dimer models. We give a non-zero vev to a
bifundamental field, which reduces the two gauge group factors under
which the bifundamental is charged to the diagonal combination. Hence,
Higgsing is nothing more than the removal of an edge from the
fundamental region of the graph, which causes two faces of the graph
to become one.

This method was used in \cite{Hanany:2005ve} to obtain the bipartite graphs corresponding to an arbitrary toric singularity, but the algorithm 
presented was computationally expensive since it was unknown how to identify the desired Higgsing in the quiver side. Using 
the duality between the quivers and brane tilings, it is straightforward to identify the edge of the bipartite graph to be removed that corresponds 
to the Higgsing of any given field in the quiver.  Thus, the relations between quiver theories under Higgsing may be easily followed on the dual 
brane tiling, avoiding any computational difficulties.

\begin{figure}[ht]
  \epsfxsize = 10cm
  \centerline{\epsfbox{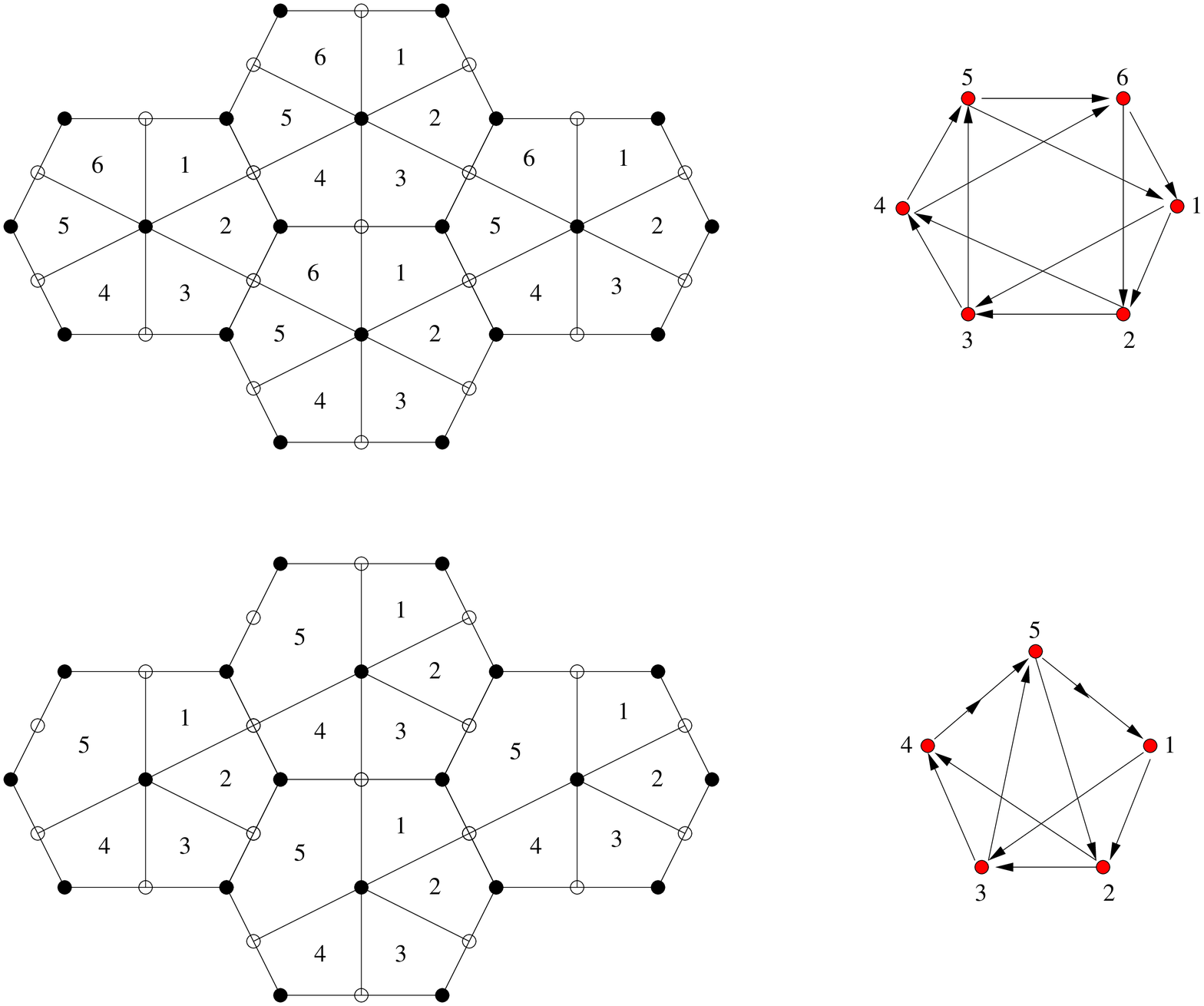}}
  \caption{Removing the edge from between faces 5 and 6 Higgses Model I of ${\bf dP}_3$ (top) to
  one of the two toric phases of ${\bf dP}_2$ (bottom).}
  \label{dp3dp2}
\end{figure}

Let us begin with Model I of ${\bf dP}_3$, since we have already studied this tiling
in a previous section. Since this model is perfectly symmetric and contains
only single bifundamental field between any two gauge groups, giving a vev
to any field should result in the same theory. This theory is ${\bf dP}_2$, which
has five nodes in its quiver. One can easily check that removing any edge from this tiling for ${\bf dP}_3$ results in the expected gauge theory. \fref{dp3dp2} illustrates this process: removing the edge between regions 5 and 6 is equivalent to removing the bifundamental between the corresponding nodes.

\begin{figure}[ht]
  \epsfxsize = 15cm
  \centerline{\epsfbox{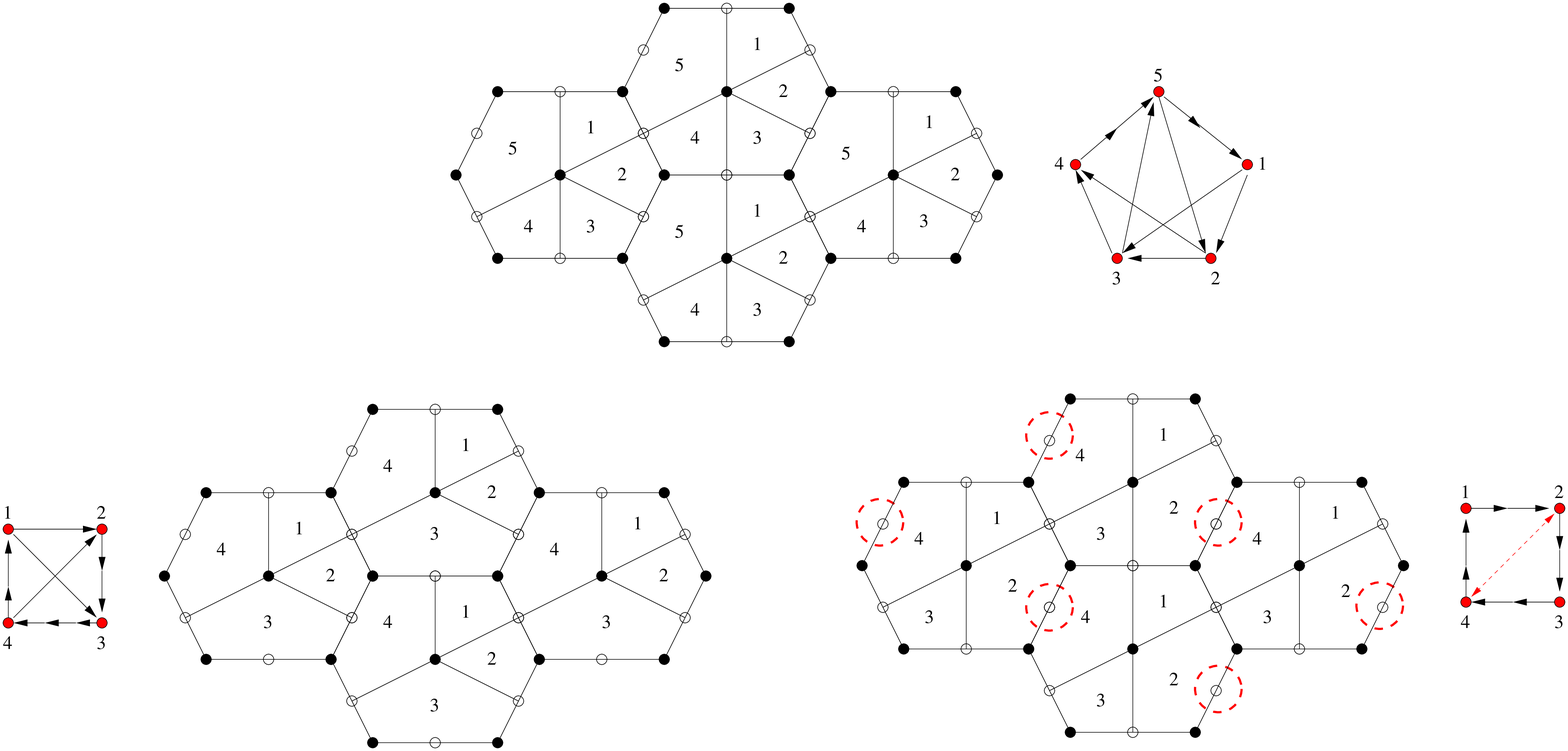}}
  \caption{The ${\bf dP}_2$ tiling (top) can be taken to either ${\bf dP}_1$ (bottom left) or ${\bf F}_0$ (bottom right), depending on which edge gets removed. In the ${\bf F}_0$ tiling, one should collapse the edge between regions 2 and 4 to a point; this corresponds to the bifundamentals on  the diagonal of the quiver.}
  \label{dp2dp1f0}
\end{figure}

The example of taking Model I of ${\bf dP}_3$ to one of the two toric phases of ${\bf dP}_2$ (called Model II in \cite{Feng:2002zw}) is particularly simple, since 
no fields acquire a mass when $X_{56}$ gets a vev. It is not any more difficult to 
see what happens when bifundamentals do become massive, as  we can see by considering the ${\bf dP}_2$ example. We know that the 
${\bf dP}_2$ theory can be Higgsed to either ${\bf dP}_1$ or ${\bf F}_0$; this corresponds to giving a vev to $X_{34}$ (or equivalently $X_{12}$ by the symmetry 
of the quiver) or $X_{23}$, respectively.  In the brane tiling, we delete the edge between regions 2 and 3 of the tiling. This puts an isolated 
node between the two regions. As per our discussion in Section \ref{section_massive_nodes}, we then simply collapse those two edges to a point, 
which corresponds to integrating out the fields $X_{35}$ and $X_{52}$. See \fref{dp2dp1f0}.

We expect from string theory that we may embed any toric quiver in an
appropriately large Abelian orbifold theory of the form $\IC^3 / \IZ_m
\times \IZ_n$. The tiling for $\IC^3 / \IZ_m \times \IZ_n$ is
hexagonal, so one expects that we can reach {\bf any} tiling by
removing edges from hexagons. This is indeed the case, as noted by
\cite{Kenyon:2003uj} and used extensively in \cite{Hanany:2005ve}.

\section{Different toric superpotentials for a given quiver}
\label{section_different_superpotentials}

Dimer methods can be used to tackle another interesting problem. Given a quiver diagram,
it is sometimes possible to construct more than one consistent toric superpotential.
Constructing the corresponding tilings shows immediately how these theories differ and
enables a straightforward computation of their toric data. 

\begin{figure}[ht]
  \epsfxsize = 4cm
  \centerline{\epsfbox{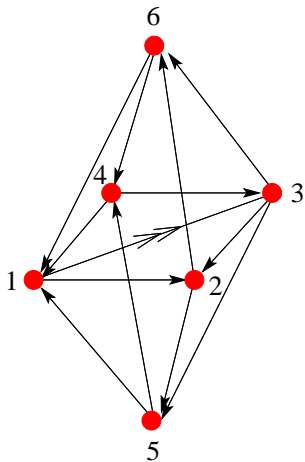}}
  \caption{Quiver diagram admitting two toric superpotentials.}
  \label{quiver_dP3_2}
\end{figure}

Let us consider a concrete example, given by the quiver diagram shown in \fref{quiver_dP3_2}.
This is the quiver for Model II of ${\bf dP}_3$ \cite{Feng:2002zw}.
This quiver has 6 gauge groups and 14 bifundamental fields. From \eref{Euler_quiver},
we see that the number of superpotential terms is $N_W=14-6=8$. 
There are two possible toric superpotentials consistent with the node symmetry group of 
the quiver. They have been considered in \cite{Feng:2002zw} and \cite{Feng:2004uq} and are 

\beq
\begin{array}{rl}
W_A & = [X_{12}X_{26}X_{61}-X_{12}X_{25}X_{51}+X_{36}X_{64}X_{43}-X_{35}X_{54}X_{43}] \\
    &+[-X_{61}X_{13}X_{36} + X_{51}Y_{13}X_{35}] + [-X_{26}X_{64}X_{41}Y_{13}X_{32} + X_{25}X_{54}X_{41}X_{13}X_{32}]
\end{array}
\label{W_dP3_II}
\eeq

\beq
\begin{array}{rl}
W_B &=Y_{13}X_{36}X_{61} + X_{13}X_{35}X_{51} - X_{61}X_{12}X_{26} - X_{43}X_{35}X_{54} \\
    &+X_{12}X_{25}X_{54}X_{41} + X_{26}X_{64}X_{43}X_{32} - X_{25}X_{51}Y_{13}X_{32} - X_{64}X_{41}X_{13}X_{36}
\end{array}
\label{W_pdP3_b}
\eeq

\begin{figure}[ht]
  \epsfxsize = 6cm
  \centerline{\epsfbox{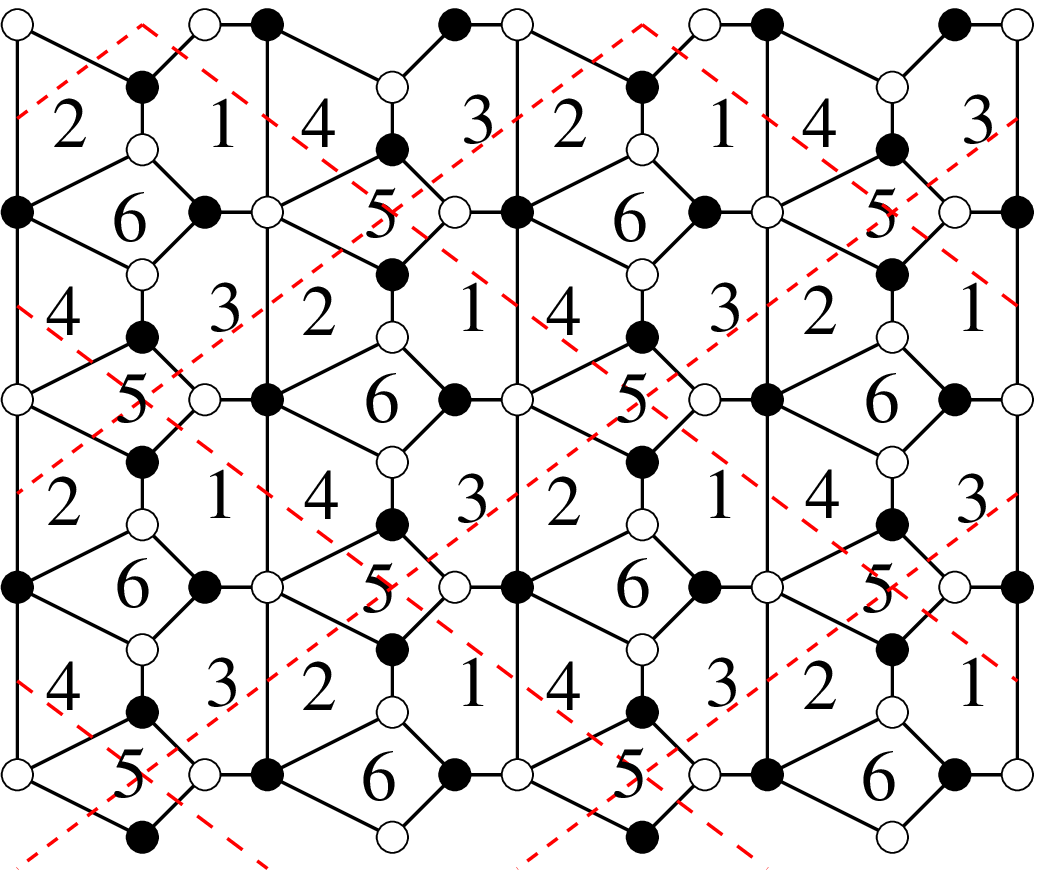}}
  \caption{Brane tiling corresponding to the quiver diagram in \fref{quiver_dP3_2} and the superpotential in
\eref{W_dP3_II}.}
  \label{dimer_dP3_II}
\end{figure}

\begin{figure}[ht]
  \epsfxsize = 5.5cm
  \centerline{\epsfbox{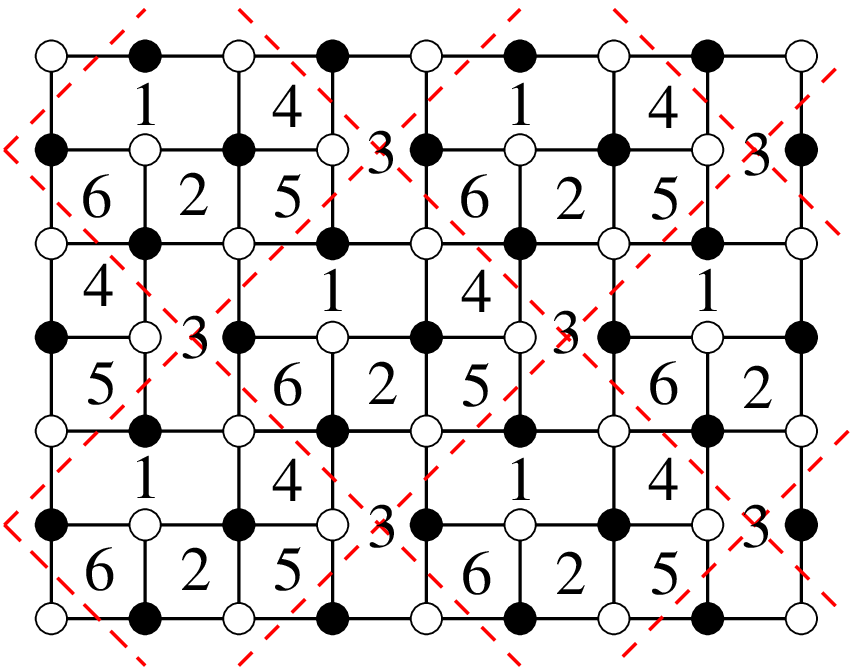}}
  \caption{Brane tiling corresponding to the quiver diagram in \fref{quiver_dP3_2} and the superpotential in
\eref{W_pdP3_b}.}
  \label{dimer_pdP3_b}
\end{figure}

\begin{figure}[ht]
  \epsfxsize = 7cm
  \centerline{\epsfbox{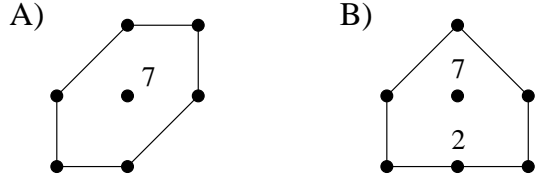}}
  \caption{Toric diagram for the quiver in \fref{quiver_dP3_2} and superpotentials $W_A$ and $W_B$}
  \label{toric_dP3_PdP3}
\end{figure}

$W_A$ corresponds to a brane tiling with six valence-3 and two valence-5 nodes. This brane
tiling is shown in \fref{dimer_dP3_II}. For $W_B$ the brane tiling has four valence-3 and
four valence-4 nodes and it is shown in \fref{dimer_pdP3_b}.
The Kasteleyn matrices for these tilings are

\beq
K_A=\left( \begin{array}{c|cccc}  & \ \ 2 \ \ & \ \ 4 \ \ & \ \ 6 \ \ & \ \ 8 \ \ \\ \hline
                             1 & 1 & z^{-1} & 0 & w \\
                             3 & -1 & 1 & 1 & 0 \\
                             5 & z+w^{-1} & w^{-1} & 1 & -z \\
                             7 & 1 & 0 & 1 & 1 \end{array} \right) \ \ \ \ \ \ \ \
K_B=\left( \begin{array}{c|cccc}  & \ \ 2 \ \ & \ \ 4 \ \ & \ \ 6 \ \ & \ \ 8 \ \ \\ \hline
                             1 & 1 & 0 & w^{-1} & w^{-1} \\
                             3 & z & 1 & 0 & w^{-1} \\
                             5 & 1 & 1 & -1 & 1 \\
                             7 & z & -1 & z & 1 \end{array} \right) 
\label{K_dP3_PdP3}
\eeq
The corresponding characteristic polynomials are

\beq
P_A(z,w)=w^{-1}z^{-1}+z^{-1}-w^{-1}+7-w+z+wz 
\label{polynomial_dP3II}
\eeq

\beq
P_B(z,w)=-w^{-2}-2w^{-1}-1-w^{-2}z+7w^{-1}z-z-w^{-1}z^2
\label{polynomial_PdP3_b}
\eeq
From \eref{polynomial_dP3II} and \eref{polynomial_PdP3_b}, we extract the toric diagrams 
shown in \fref{toric_dP3_PdP3}.

Thus we see that $W_A$ leads to Model II of ${\bf dP}_3$ (the multiplicities of GLSM fields are in agreement with
the ones derived in \cite{Feng:2002zw}) while $W_B$ leads to a non-generic blow-up of $\IC\IP^2$ at three
points, denoted $PdP_{3b}$ in \cite{Feng:2004uq}.

\section{Examples}
\label{sec:examples}

Here we present the brane tiling configurations for several interesting gauge theories. Many of them can be obtained using the Seiberg duality and 
partial resolution ideas discussed in previous sections. When doing so, we generate data on GLSM multiplicities for all these models.

\subsection{Del Pezzo 2}

There are two toric phases for ${\bf dP}_2$. Their corresponding quivers and superpotentials can be found in \cite{Feng:2002zw}. We now construct their corresponding brane tilings.

\subsection*{Model I}

\begin{figure}[ht]
  \epsfxsize = 5.5cm
  \centerline{\epsfbox{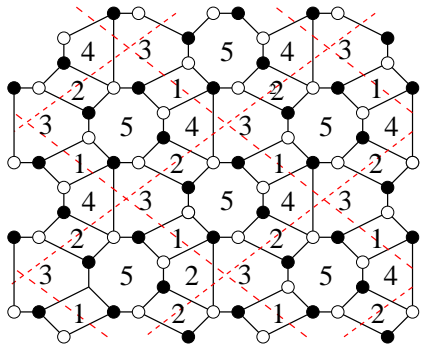}}
  \caption{Brane tiling for Model I of ${\bf dP}_2$.}
  \label{dimer_dP2_I}
\end{figure}

The brane tiling for this model is shown in \fref{dimer_dP2_I}.
The Kasteleyn matrix is

\beq
K=\left( \begin{array}{cccc}   \ 1 \ & \ w^{-1} \ & \ w^{-1} z^{-1} \ & \ 1 \ \\ 
                             1 & 1 & -z^{-1} & 0 \\
			     0 & 1 & -1 & -w \\
                             z & 0 &  1 & 1 \end{array} \right)
\eeq
leading to

\beq
P(z,w)=w^{-1}z^{-1}+z^{-1}+w^{-1}-6+w+z
\label{pol_dP2_1}
\eeq

\subsection*{Model II}

\begin{figure}[ht]
  \epsfxsize = 7.0cm
  \centerline{\epsfbox{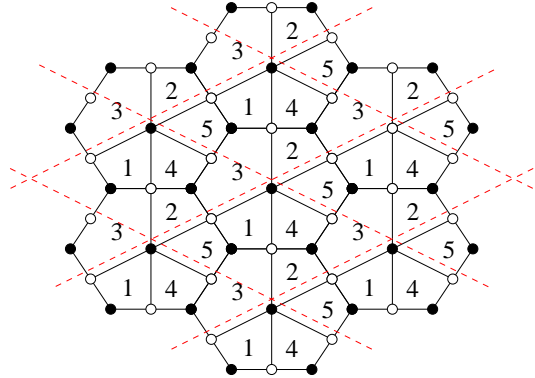}}
  \caption{Brane tiling for Model II of ${\bf dP}_2$.}
  \label{dimer_dP2_II}
\end{figure}

The tiling for this model was obtained in Section \ref{section_partial_resolution} by means 
of partial resolution. We show it again in \fref{dimer_dP2_II}.
The corresponding Kasteleyn matrix is

\beq
K=\left( \begin{array}{ccc}   \ 1-z^{-1} \ & \ w \ & \ 1 \ \\ 
                             1 & 1 & z  \\
			     -1+w^{-1}z^{-1} & 1 & 1 \end{array} \right)
\eeq
which leads to the following characteristic polynomial
\beq
P(z,w)=w^{-1}z^{-1}-z^{-1}+5-w-z-wz
\label{pol_dP2_2}
\eeq
From \eref{pol_dP2_1} and \eref{pol_dP2_2} we can determine the toric diagrams along with the GLSM multiplicities, 
which are in agreement with the results in \cite{Feng:2002zw}.

\subsection*{Del Pezzo 3}

There are four toric phases for ${\bf dP}_3$. We refer the reader to \cite{Feng:2002zw} for their quivers
and superpotentials. We have already presented the tiling for Model I in \fref{dimer_quiver_dP3_1}. 
Its Kasteleyn matrix and characteristic polynomial are written in \eref{K_dP3_1} and \eref{pol_dP3_1}.
\fref{dimer_dP3_II} shows the tiling for Model II. Its Kasteleyn matrix and characteristic polynomial are presented
in \eref{K_dP3_PdP3} and \eref{polynomial_dP3II}. We now proceed with the construction of the brane tilings
for Models III and IV.

\bigskip

\subsection*{Model III}

\begin{figure}[ht]
\epsfxsize = 5.5cm
  \centerline{\epsfbox{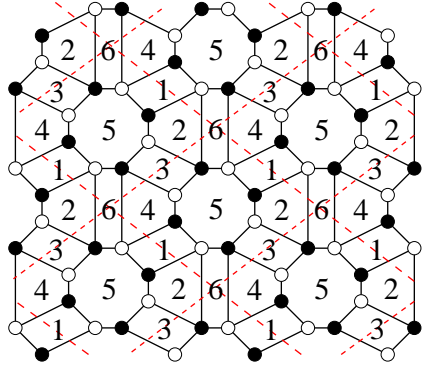}}
  \caption{Brane tiling for Model III of ${\bf dP}_3$.}
  \label{dimer_dP3_III}
\end{figure}

We show the brane tiling in \fref{dimer_dP3_III}.
The Kasteleyn matrix is

\beq
K=\left( \begin{array}{c|cccc}   
                             & \ \ 2 \ \ & \ \ 4 \ \ & \ \ 6 \ \ & \ \ 8 \ \ \\ \hline
                             1 &  1  &  w^{-1}  &  w^{-1} z^{-1}  &  1  \\ 
                             3 & 1 & 1 & -z^{-1} & 0 \\
			     5 & wz & 1 & -1 & -w \\
                             7 & z & 0 &  1 & 1 \end{array} \right)
\eeq
from which we compute the determinant

\beq
P(z,w)=w^{-1}z^{-1}+z^{-1}-w^{-1}-8+w+z+wz
\eeq

This corresponds to the toric diagram of ${\bf dP}_3$ with multiplicity 8 for the central point. 
This result agrees with the Forward Algorithm computations in \cite{Feng:2002zw}.

\subsection*{Model IV}

\begin{figure}[h]
  \epsfxsize = 6.5cm
  \centerline{\epsfbox{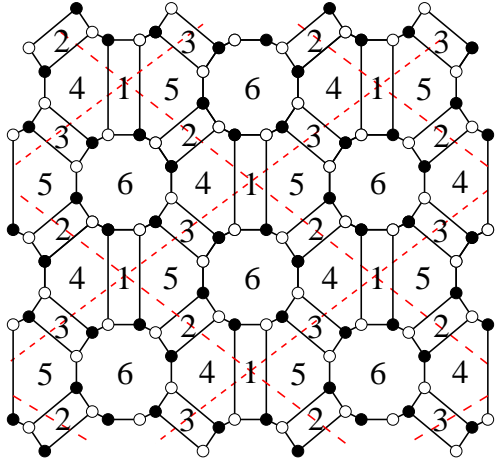}}
  \caption{Brane tiling for Model IV of ${\bf dP}_3$.}
  \label{dimer_dP3_IV}
\end{figure}

\fref{dimer_dP3_IV} shows the brane tiling for this theory.

The Kasteleyn matrix is given by

\beq
K=\left( \begin{array}{c|cccccc}  & \ \ 2 \ \ & \ \ 4 \ \ & \ \ 6 \ \ & \ \ 8 \ \ & \ \ 10 \ \ & \ \ 12 \ \  \\ \hline
                             1 & 1 & 0 & 0 & -wz & 0 & -1 \\
                             3 & 1 & 1 & 0 & 0 & z & 0 \\
                             5 & 0 & -1 & 1 & 0 & 0 & -w^{-1} \\
                             7 & w^{-1} z^{-1} & 0 & 1 & 1 & 0 & 0 \\
                             9 & 0 & -z^{-1} & 0 & -1 & 1 & 0 \\
                            11 & 0 & 0 & w & 0 & 1 & 1 \end{array} \right)
\eeq
and the characteristic polynomial is

\beq
P(z,w)=-w^{-1}z^{-1}-z^{-1}-w^{-1}+11-w-z-wz
\eeq

Once again, this corresponds to the toric diagram of ${\bf dP}_3$. In this case, the multiplicity of the central point is 11,
in agreement with the computations in \cite{Feng:2002zw}.

\vfill
\pagebreak
\subsection{Pseudo del Pezzo 5}

\label{section_PdP_5}

\begin{figure}[ht]
  \epsfxsize = 15cm
  \centerline{\epsfbox{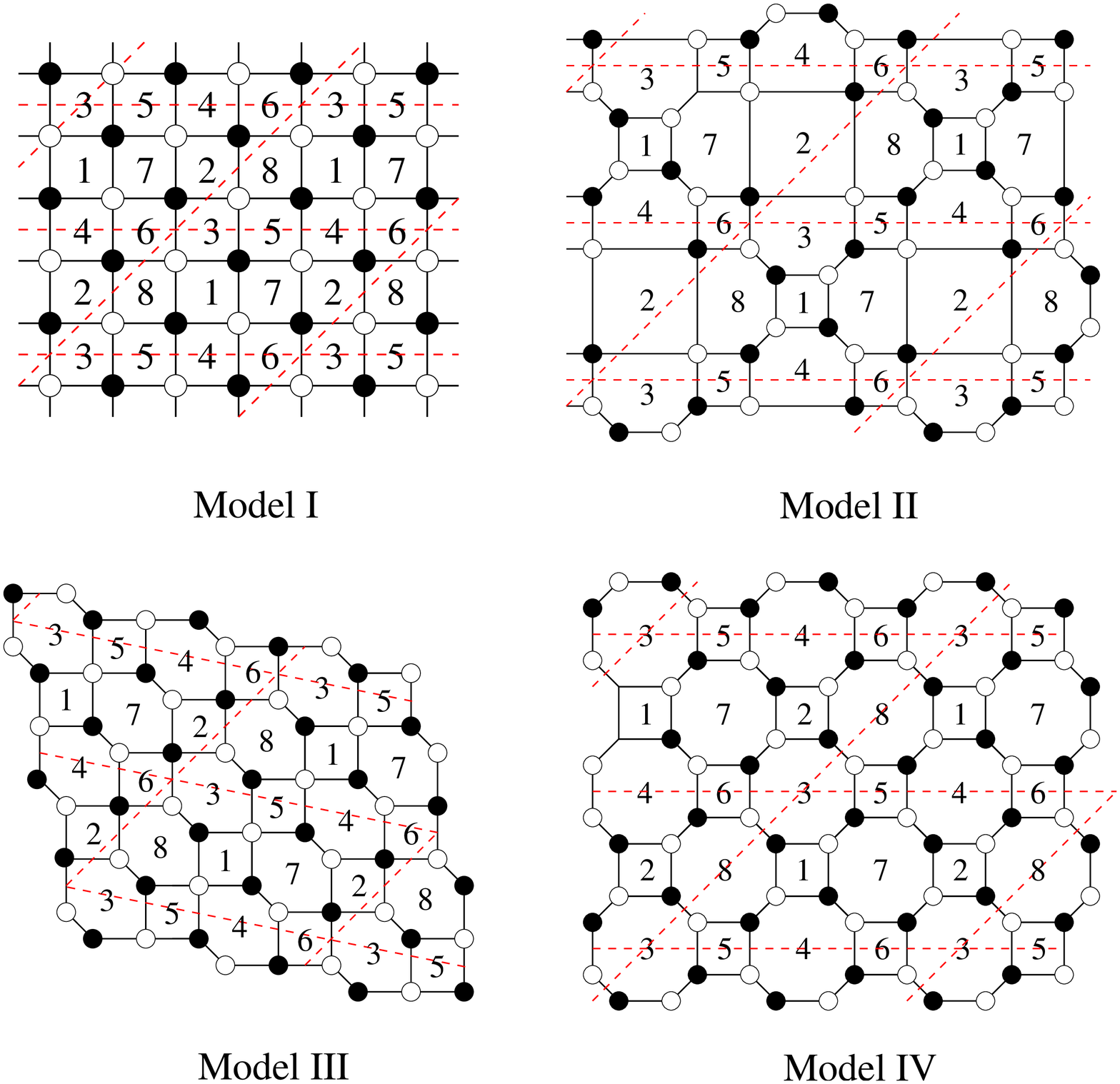}}
  \caption{Brane tilings for the four toric phases of $PdP_5$.}
  \label{dimers_PdP5}
\end{figure}

We now consider a complex cone over non-generic, toric blow-up of $\IC\IP^2$ at 
five points. The geometry corresponds to a $\IZ_2 \times \IZ_2$ orbifold of the conifold
and was dubbed $PdP_5$ in \cite{Feng:2002fv}, where the corresponding gauge theories were studied.
There are four toric phases for this geometry. We refer the reader to \cite{Feng:2002fv} for the quivers
and superpotentials. The brane tilings for these four phases are shown in \fref{dimers_PdP5}

{\small
\beq
\hspace{-1.5cm}
\begin{array}{ccc}

K_{I}=\left( \begin{array}{c|cccc}   
                             & \ \ 2 \ \ & \ \ 4 \ \ & \ \ 6 \ \ & \ \ 8 \ \ \\ \hline
                             1 &  -1  &  -w  & 1  &  wz  \\ 
                             3 &  -1  &  -1  & z & 1 \\
			           5 & z^{-1} & w & 1 & w \\
                             7 & 1 & z^{-1} &  1 & 1 \end{array} \right) & \ \ \ \ \ \ &
K_{II}=\left( \begin{array}{c|cccccc}   
                             & \ \ 2 \ \ & \ \ 4 \ \ & \ \ 6 \ \ & \ \ 8 \ \ & \ \ 10 \ \ & \ \ 12 \ \ \\ \hline
                             1 &  -1  &  0  &  -w  &  0 & 0 & wz  \\ 
                             3 & -1 & 1 & 0 & 0 & 1 & 0 \\
			           5 & 0 & -1 & -1 & z & 0 & 1 \\ 
                             7 & 0 & z^{-1} & w & 1 & 0 & w \\ 
                             9 & 1 & 0 & 0 & 1 & 1 & 0 \\
                             11 & 0 & 0 & z^{-1} & 0 & -1 & 1\end{array} \right)\\ \\ \\
K_{III}=\left( \begin{array}{c|cccccc}   
                             & \ \ 2 \ \ & \ \ 4 \ \ & \ \ 6 \ \ & \ \ 8 \ \ & \ \ 10 \ \ & \ \ 12 \ \ \\ \hline
                             1 &  1  &  0  &  w  &  0 & 0 & -wz  \\ 
                             3 & 1 & 1 & 0 & -z & -1 & 0 \\
			           5 & 0 & -1 & 1 & 0 & 0 & -1 \\ 
                             7 & 0 & 0 & -w & -1 & 0 & -w \\ 
                             9 & -1 & -z^{-1} & 0 & -1 & -1 & 0 \\
                             11 & 0 & 0 & -z^{-1} & 0 & 1 & -1 \end{array} \right) & \ \ \ \ \ \ &
K_{IV}=\left( \begin{array}{c|cccccccc}   
                             & \ \ 2 \ \ & \ \ 4 \ \ & \ \ 6 \ \ & \ \ 8 \ \ & \ \ 10 \ \ & \ \ 12 \ \ & \ \ 14 \ \ & \ \ 16 \ \ \\ \hline
                             1 &  1  &  0  &  0  &  w & 0 & 0 & wz & 0  \\ 
                             3 & -1 & 1 & 0 & 0 & 0 & 1 & 0 & 0 \\
			           5 & 0 & 1 & 1 & 0 & z & 0 & 0 & 0 \\ 
                             7 & 0 & 0 & -1 & 1 & 0 & 0 & 0 & 1 \\ 
                             9 & 0 & z^{-1} & 0 & 0 & -1 & 0 & 0 & w \\
                             11 & 1 & 0 & 0 & 0 & 1 & 1 & 0 & 0 \\
                             13 & 0 & 0 & 0 & z^{-1} & 0 & 1 & -1 & 0 \\
                             15 & 0 & 0 & 1 & 0 & 0 & 0 & 1 & 1  \end{array} \right) 
\end{array}
\eeq}

From these matrices, we compute the corresponding characteristic polynomials

\beq
\begin{array}{rl}
P_I(z,w)=P_{III}(z,w)= & z^{-2}+2 z^{-1}+2w z^{-1}+1-12w+w^2+2wz+2w^2z+w^2z^2 \\
P_{II}(z,w)=& z^{-2}+2 z^{-1}+2w z^{-1}+1-14w+w^2+2wz+2w^2z+w^2z^2 \\
P_{IV}(z,w)=& z^{-2}-2 z^{-1}-2w z^{-1}+1-21w+w^2-2wz-2w^2z+w^2z^2 
\end{array}
\eeq

\begin{figure}[h]
  \epsfxsize = 11cm
  \centerline{\epsfbox{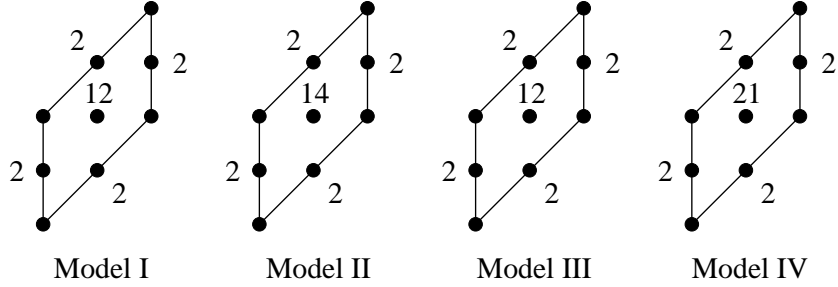}}
  \caption{Toric diagrams with multiplicities for the four toric phases of $PdP_5$. We observe that the GLSM 
multiplicities are the same for Models I and III.}
  \label{toric_PdP5}
\end{figure}

Remarkably, although $K_I$ and $K_{III}$ are different matrices with different dimensions,
their characteristic polynomials turn out to be identical. This is a counterexample
to the conjecture that GLSM multiplicities are in one to one correspondence with the dual phases
of the gauge theory. Different phases can indeed lead to exactly the same multiplicities.
We present the toric diagrams with multiplicities in \fref{toric_PdP5}.

\subsection{Tilings for infinite families of gauge theories}
\label{sec:infquivers}

One of the problems for which dimer methods show their full power is in the determination
of dual geometries for infinite families of gauge theories. Infinite sets of quiver
theories have recently been constructed in \cite{Benvenuti:2004dy} and \cite{Hanany:2005hq}.
On one hand, we have already discussed that the application of the Forward Algorithm 
to large quivers becomes computationally prohibitive. In addition, it is obviously 
impossible to apply the Forward Algorithm to an infinite number of theories.
Hence, the determination of gauge theories dual to an infinite number of geometries
usually involve indirect evidence such as: (un)higgsing, global symmetries, computation of 
R-charges and central charges and comparison to volumes in the underlying geometry \cite{Benvenuti:2004dy,Hanany:2005hq}.

\subsubsection{$Y^{p,q}$ tilings}

\newcommand{\n}{\noindent}

\newcommand{\baq}{\begin{eqnarray}}
\newcommand{\eaq}{\end{eqnarray}}
\newcommand{\nn}{\nonumber\\}

\newcommand{\pp}{\partial}
\newcommand{\nl}{\newline}

\def\ph{{\phantom{1}}}        
\def\phb{{\phantom{\Big|}}}   

\def\itm{\mathrm{i.m.}} 

\def\even{\mathrm{even}}
\def\odd{\mathrm{odd}}

\def\ccdot{\raisebox{1mm}{${}_\bullet$}}
\def\std{\mathrm{std}}
\def\Fix{\mathrm{Fix}}
\def\res{\!\!\upharpoonright\!\!}  
\def\dag{\dagger}          
\def\transp{t}             

\newcommand{\placeit}[3]{\raisebox{#2}[0pt][0pt]{\makebox[0pt][l]{\hspace{#1}#3}}}

\def\dirlim{\placeit{1pt}{-4.7pt}{$\lra$}\lim\,}
\def\invlim{\placeit{1pt}{-4.7pt}{$\lla$}\lim\,}

\def\lmod{\textsf{-mod}}
\def\lind{\textsf{-ind}}
\def\lproj{\textsf{-proj}}
\def\linj{\textsf{-inj}}
\def\rmod{\textsf{mod-}}
\def\rind{\textsf{ind-}}
\def\rproj{\textsf{proj-}}
\def\rinj{\textsf{inj-}}
\def\liegr{\textsf{liegr}}
\def\liealg{\textsf{liealg}}
\def\affalg{\textsf{affalg}}
\def\projalg{\textsf{projalg}}
\def\ctop{\textsf{top}}
\def\ltop{\textsf{-top}}
\def\repr{\textsf{-repr}}
\def\quiv{\textsf{quiv}}

\def\SS{\mathfrak{S}}

\def\DD{\mathcal{D}}  
\def\W{\mathfrak{W}}  
\def\C{\mathcal{C}}   
\def\WW{\mathcal{W}}  
\def\XX{\mathcal{X}}  
\def\HH{\mathcal{H}}  

\def\AA{\mathcal{A}}
\def\II{\mathcal{I}}
\def\JJ{\mathcal{J}}
\def\KK{\mathcal{K}}
\def\Nn{\mathcal{N}}
\def\UU{\mathcal{U}}
\def\VV{\mathcal{V}}
\def\YY{\mathcal{Y}}

\def\gg{\mathfrak{g}}
\def\hh{\mathfrak{h}}
\def\zz{\mathfrak{z}} 

\def\diag{\mathrm{diag}}
\def\Ad{\mathrm{Ad}}
\def\ad{\mathrm{ad}}
\def\rank{\mathrm{rank}}
\def\corank{\mathrm{corank}}
\def\rk{\mathrm{rk}}
\def\Aut{\mathrm{Aut}}
\def\pr{\mathrm{pr}}
\def\Tp{\mathrm{Tp}}
\def\tp{\mathrm{tp}}
\def\deg{\mathrm{deg}}
\def\dim{\mathrm{dim}}
\def\Stab{\mathrm{Stab}}
\def\pt{\mathrm{pt}}
\def\Gr{\mathrm{Gr}}

\def\OO{\mathcal{O}}
\def\SS{\mathfrak{S}}

\def\GL{\mathrm{GL}}
\def\SL{\mathrm{SL}}
\def\U{\mathrm{U}}
\def\SU{\mathrm{SU}}
\def\O{\mathrm{O}}
\def\SO{\mathrm{SO}}
\def\Sp{\mathrm{Sp}}
\def\PGL{\mathrm{PGL}}

\def\gl{\mathfrak{gl}}
\def\sl{\mathfrak{sl}}
\def\u{\mathfrak{u}}
\def\su{\mathfrak{su}}
\def\o{\mathfrak{o}}
\def\so{\mathfrak{so}}
\def\sp{\mathfrak{sp}}

\def\la{\leftarrow}
\def\ra{\rightarrow}
\def\lla{\longleftarrow}
\def\lra{\longrightarrow}
\def\acts{\!\curvearrowright\!}
\def\racts{\!\curvearrowleft\!}
\def\embed{\hookrightarrow}
\def\isom{\cong}

\def\lisom{\placeit{2.7mm}{1.5mm}{$\sim$}\longleftarrow}
\def\risom{\placeit{2.5mm}{1.5mm}{$\sim$}\longrightarrow}
\def\llisom{\placeit{3.5mm}{1.5mm}{$\sim$}\xleftarrow   {\;\;\,\,\;\;\;}}
\def\lrisom{\placeit{3.2mm}{1.5mm}{$\sim$}\xrightarrow{\;\;\,\,\;\;\;}}
\def\rcup{\xrightarrow{\;\,\smallsmile\;\;}}
\def\rcross{\xrightarrow{\;\,\times;\;}}

\def\phi{\varphi}
\def\epsilon{\varepsilon}
\def\Nabla{\nabla}

\def\NN{\mathbb{N}}
\def\ZZ{\mathbb{Z}}
\def\QQ{\mathbb{Q}}
\def\RR{\mathbb{R}}
\def\CC{\mathbb{C}}

\def\Aff{\mathbb{A}}
\def\Proj{\mathbb{P}}

\def\tr{\mathrm{tr}}
\def\Diff{\mathrm{Diff}}
\def\Hom{\mathrm{Hom}}
\def\HOM{\mathrm{HOM}}
\def\id{\mathrm{id}}
\def\Sym{\mathrm{Sym}}
\def\deg{\mathrm{deg}}
\def\dim{\mathrm{dim}}
\def\codim{\mathrm{codim}}
\def\im{\mathrm{Im}}
\def\ker{\mathrm{Ker}}

\def\Mat{\mathrm{Mat}}

\def\vacuum{|0\rangle}

\begin{figure}[ht]
  \epsfxsize = 6cm
  \centerline{\epsfbox{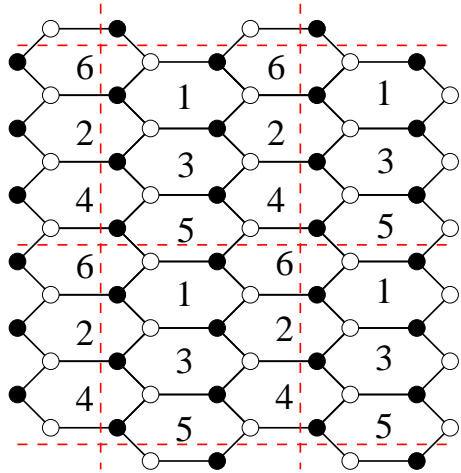}}
  \caption{Brane tiling for $Y^{3,3}$}
  \label{Y33}
\end{figure}

Let us discuss now how the $Y^{p,q}$ theories \cite{Benvenuti:2004dy} appear in the brane tiling picture. A simple way to construct the $Y^{p,q}$'s is to start 
with $Y^{p,q=p}$ and decrease $q$ by introducing ``impurities'' into the quiver \cite{Benvenuti:2004wx}. This procedure can be similarly carried 
out with tilings. Since $Y^{p,p}$ is the base of the orbifold $\CC^3/\ZZ_{2p}$, it corresponds to the hexagonal graph with a 
fundamental cell containing $2 \times p$ hexagons. This is shown in \fref{Y33} for $Y^{3,3}$.

\begin{figure}[ht]
  \epsfxsize = 6cm
  \centerline{\epsfbox{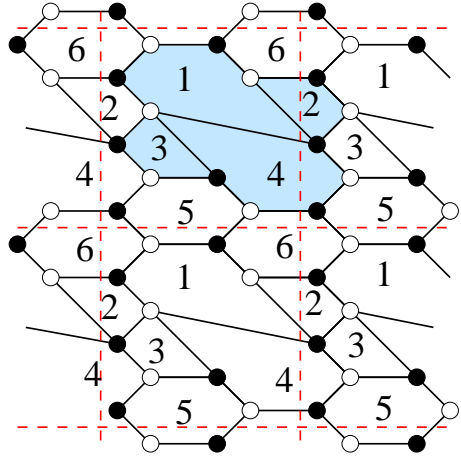}}
  \caption{Brane tiling for $Y^{3,2}$. The impurity is the blue area.}
  \label{Y32}
\end{figure}

\begin{figure}[ht]
  \epsfxsize = 8cm
  \centerline{\epsfbox{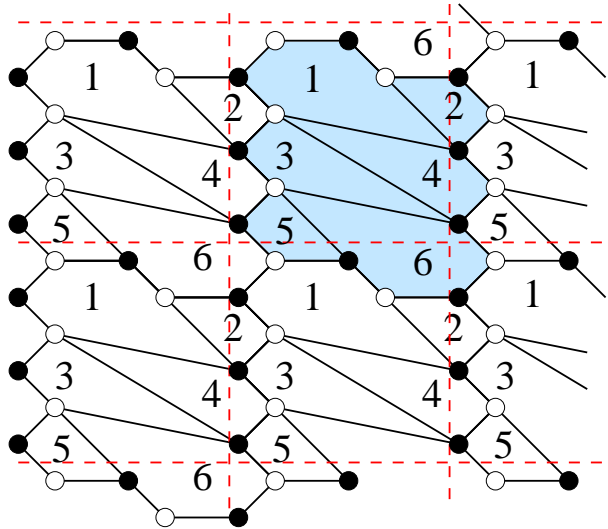}}
  \caption{Brane tiling for $Y^{3,1}$.}
  \label{Y31}
\end{figure}

Let us put now a single impurity into the tiling. The impurity covers four hexagons, and is indicated in blue in \fref{Y32}.
Two disjoint single impurities can be generated by adding an identical shaded region
into the tiling, separated from the first one by some hexagonal faces. 
For $Y^{3,1}$ this is not possible because the fundamental cell
consists of only six hexagons, whereas two separated single impurities 
would cover eight of them. Instead, we can consider the case in which the two
impurities are adjacent. This corresponds to a similar impurity graph, 
which is shown in \fref{Y31}.

One can continue adding impurities and discover the simplicity of the Kasteleyn matrix 
for $Y^{p,q}$. It contains elements only in the diagonal and its neighbors and in the corners. 
It can be written down immediately, without actually drawing the corresponding 
brane tiling. One starts with the following $2p \times 2p$ Kasteleyn matrix
for $Y^{p,p}$.

\beq
K = \left( \begin{array}{cccccccc} 
1 & 1 & 0 & 0 & \hdotsfor{2} & 0& z^{-1} \\
1 & w & 1 & 0 & \hdotsfor{3}  & 0\\
0 & 1 & 1 & 1 & 0 & \hdotsfor{2} & 0\\
0 & 0 & 1 & w & 1 & 0 & \ldots  & 0\\
& \vdots & & & & \ddots & & \\
0 & \hdotsfor{3} & 0 & 1 & 1 & 1 \\
z & 0 & \hdotsfor{3} & 0 & 1 & w
\end{array} \right)
\eeq
We see that the elements around the diagonal consist of the alternating ``codons'':
\baq
  A_1&:=&(1,1,1) \\
  A_2&:=&(1,w,1)
\eaq
We define three other codons 
\baq
  S&:=&(1,w,w) \\
  I&:=&(1,-1+w,w) \\
  E&:=&(1,-1+w,1)
\eaq

\begin{figure}[ht]
  \epsfxsize = 1.5cm
  \centerline{\epsfbox{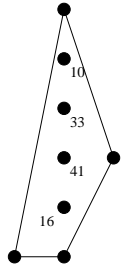}}
  \caption{Toric diagram of a phase of $Y^{5,3}$}
  \label{Y53td}
\end{figure}

Placing impurities into the quiver means changing the $A_1, A_2, A_1, A_2, \dots$
sequence in the matrix. For $n$ single impurities we get $Y^{p,p-n}$ and the Kasteleyn
matrix gets smaller, it is now a $(2p-n) \times (2p-n)$ matrix.
We change the sequence of the codons as the following. In the second row we put
$S$ (start codon), then $n-1$ times the $I$ (iteration codon), and we close it
with $E$ (end codon). Then we continue the series with $A_1, A_2, A_1, A_2, \dots$
until the end of the matrix. As an example, we present the Kasteleyn matrix for
$Y^{5,3}$ (i.e.~$n=2$)
\beq
K = \left( \begin{array}{c|ccccccccccccc} 
A_1 && 1 && 1 & 0 & 0 & 0 && 0 && 0 && z^{-1} \\
S && 1 && w & w & 0 & 0 && 0 && 0 && 0 \\
I && 0 && 1 & -1+w & w & 0 && 0 && 0 && 0 \\
E && 0 && 0 & 1 & -1+w & 1 && 0 && 0 && 0 \\
A_1 && 0 && 0 & 0 & 1 & 1 && 1 && 0 && 0 \\
A_2 && 0 && 0 & 0 & 0 & 1 && w && 1 && 0 \\
A_1 && 0 && 0 & 0 & 0 & 0 && 1 && 1 && 1 \\
A_2 && z && 0 & 0 & 0 & 0 && 0 && 1 && w \\
\end{array} \right)
\eeq
The determinant of the Kasteleyn matrix is then
\beq
  P(w,z) = -1 + 16w - 41w^2 + 33w^3 - 10w^4 + w^5 - z^{-1} - w^2 z,
\eeq
and the toric diagam (with GLSM multiplicities) is given in \fref{Y53td}.

We note that the above rules for constructing the Kasteleyn matrix produce
the toric diagrams for all $Y^{p,q}$ with $p>q>0$. To check this, we can see that the correct
monomials appear in the determinant. First, the only powers of $z$ that appear
in $\det K$ are -1,0, and 1. Terms of the form $z^0 w^k$ appear for all $k=0,\dots,p$;
these come from the diagonal. Second, there is a term $z^{-1}w^0$ that comes from
the lower off-diagonal. Finally, the term $zw^n$, where $n=p-q$ is the number
of single impurities, comes from the upper off-diagonal and gets contributions
from only the $S$ and $I$ codons. Thus, we have shown that 
the moduli spaces of the $Y^{p,q}$ quivers reproduce the correct toric
geometries.  

\begin{figure}[ht]
  \epsfxsize = 1.5cm
  \centerline{\epsfbox{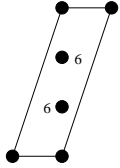}}
  \caption{Toric diagram of a phase of $Y^{3,0}$ with three single impurities.}
  \label{Y30td}
\end{figure}

For $Y^{p,0}$ the matrix gets too small and there is not enough space for
$A_1$ and $A_2$. The Kasteleyn matrix consists of only $I$ codons:
\beq
K = \left( \begin{array}{cccccccc} 
-1+w & w & 0 & 0 & \hdotsfor{3} & z^{-1} \\
1 & -1+w & w & 0 & \hdotsfor{3} & 0\\
0 & 1 & -1+w & w & 0 & \hdotsfor{2} & 0\\
0 & 0 & 1 & -1+w & w & 0 & \ldots  & 0\\
& \vdots & & & & \ddots & & \\
0 & \hdotsfor{3} & 0 & 1 & -1+w & w \\
wz & 0 & \hdotsfor{3} & 0 & 1 & -1+w
\end{array} \right)
\eeq
For example, the Kasteleyn matrix of $Y^{3,0}$ is:
\beq
K = \left( \begin{array}{ccc} 
-1+w & w & z^{-1} \\
1 & -1+w & w \\
wz & 1 & -1+w 
\end{array} \right)
\eeq
with determinant
\beq
  P(w,z) = -1 + 6w - 6w^2 + w^3 + z^{-1} + w^3 z
\eeq
and the toric diagram of \fref{Y30td}.

These Kasteleyn matrices give the toric diagrams of a certain phase of
the theories (the one with only single impurities, all of them together).  
Other phases can be obtained by performing Seiberg duality transformations.
As discussed in section \ref{sec:seiberg} this may be efficiently
implemented on a computer and used to enumerate the toric phases of
the theory, together with the duality graph showing the interconnections between phases.

\subsubsection{$Y^{3,1}$ with double impurity}

\begin{figure}[ht]
  \epsfxsize = 16cm
  \centerline{\epsfbox{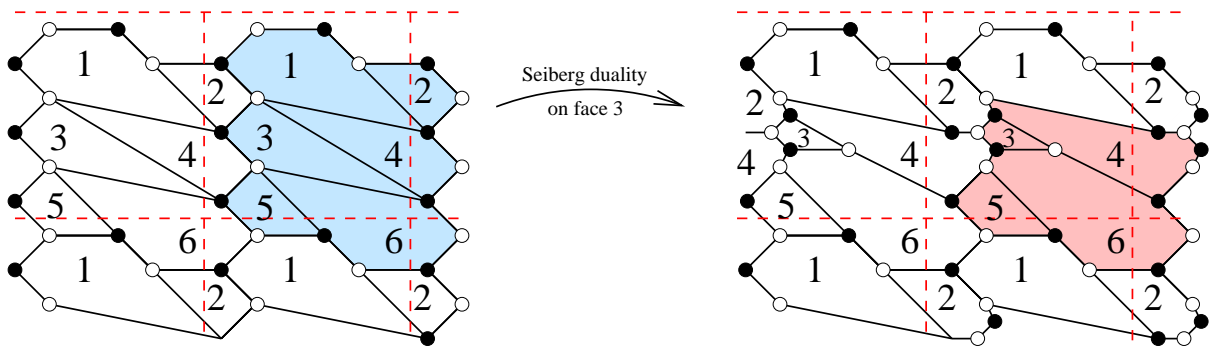}}
  \caption{Dualizing face $3$ in $Y^{3,1}$ with two single impurities. In resulting tiling, we indicate the double
impurity in pink.}
  \label{Y31_double_imp_dual}
\end{figure}

\begin{figure}[ht]
  \epsfxsize = 8cm
  \centerline{\epsfbox{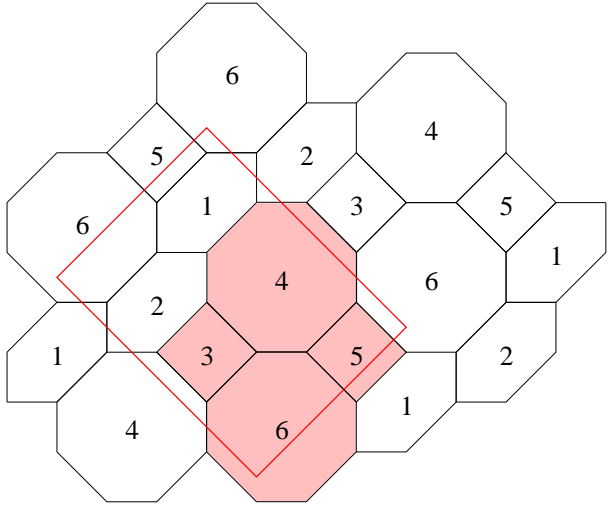}}
  \caption{The double impurity in $Y^{3,1}$}
  \label{Y31d}
\end{figure}

\begin{figure}[ht]
  \epsfxsize = 5.5cm
  \centerline{\epsfbox{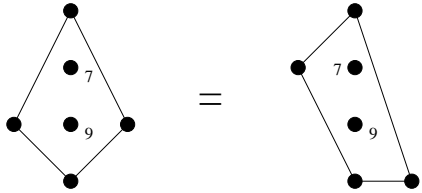}}
  \caption{Toric diagram for $Y^{3,1}$ in the double impurity phase}
  \label{Y31dtd}
\end{figure}

In \cite{Benvenuti:2004wx}, it was shown that all toric phases of $Y^{p,q}$ theories
can be constructed by adding single and double impurities to the $\IC^3/\IZ_{2p}$
quiver. Double impurities arise when Seiberg duality makes two single impurities ``collide''.
As an example of Seiberg duality, we now study the double impurity phase
of $Y^{3,1}$. This phase can be obtained by dualizing face $3$ (see \fref{Y31_double_imp_dual}).
The resulting graph can be deformed to the more symmetric form which is shown in
\fref{Y31d}. The determinant of the Kasteleyn matrix again gives the $P(w,z)$ polynomial,
from which we get the toric diagram (\fref{Y31dtd}).

\beq
K = \left( \begin{array}{cccccc} 
1 & 1 & 0 & 0 & 0 & z^{-1} \\
w^{-1} & 1 & w^{-1} & 0 & 0 & 0 \\
0 & 1 & 1 & 1 & 0 & 0 \\
0 & 0 & w^{-1} & 1 & 1 & 0 \\
0 & 0 & 0 & 1 & -1 & -1 \\
z^{-1} & 0 & 0 & 0 & w & 1 \\
\end{array} \right)
\eeq

\beq
  P(w,z) = -7 - w^{-2} + 9w^{-1} + w + w^{-1} z^{-1} + z w^{-1}
\eeq

All the multiplicity results in this and previous section agree with the ones
derived using the Forward Algorithm in \cite{Dunn:2005a}.

\subsubsection{$X^{p,q}$ tilings}

We now describe the brane tilings for the $X^{p,q}$ spaces 
constructed in \cite{Hanany:2005hq}. Recall that these spaces are defined by the property
that an $X^{p,q}$ theory can be Higgsed to both $Y^{p,q}$ and $Y^{p,q-1}$. Constructing
the brane tilings for the $X^{p,q}$ is quite straightforward, but it will be convenient
for our purposes to use a slightly modified (but entirely equivalent) description of
the $Y^{p,q}$ spaces from the one used in the previous section. 

\begin{figure}[ht]
  \epsfxsize = 15cm
  \centerline{\epsfbox{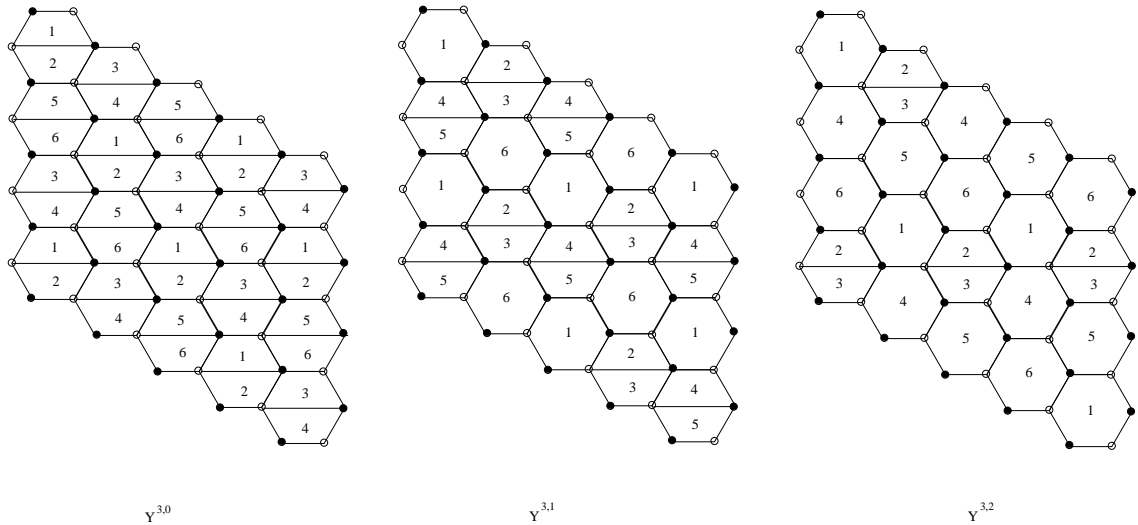}}
  \caption{Brane tilings for $Y^{3,q}$.}
  \label{y3qhex}
\end{figure}

We use the following description of $Y^{p,q}$, with $p-q$ single impurities. 
For this tiling, we need $2(p-q)$ quadrilaterals and $2q$ hexagons. We build the
quadrilaterals by starting with a hexagonal grid, and drawing lines through the center of
a given hexagon, connecting opposite vertices. This divides the hexagon into two quadrilaterals.
A given $Y^{p,q}$ is then given by placing these hexagons and divided
hexagons along a single diagonal such that the divided hexagons are separated from
each other by an even number of non-divided hexagons; this is simply the requirement
that the single impurities be separated from each other by an odd number of doublets. 
For examples of this construction, see \fref{y3qhex}.

\begin{figure}[ht]
  \epsfxsize = 6cm
  \centerline{\epsfbox{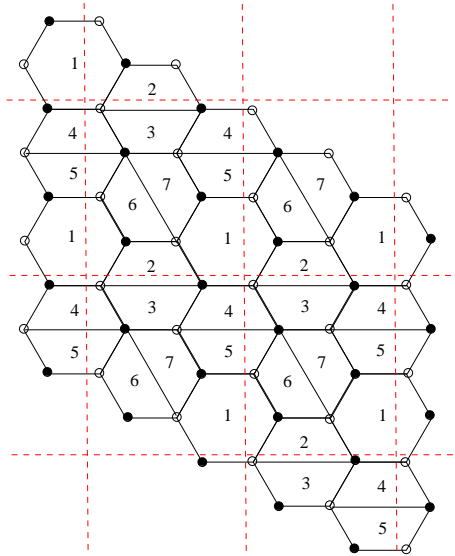}}
  \caption{A brane tiling for $X^{3,1}$.}
  \label{x31}
\end{figure}

Constructing the $X^{p,q}$ brane tilings is now straightforward. We give the example
of $X^{3,1}$ below; the other $X^{p,q}$ tilings work similarly. To build the tilings,
simply insert diagonal lines in hexagons such that removing
the line from the $X^{p,q}$ tiling gives the $Y^{p,q}$ to which it descends. 
This diagonal line should always share a node with one of the horizontal
lines subdividing a hexagon in two; this is what allows
one to blow down to $Y^{p,q-1}$ as well as $Y^{p,q}$. In \fref{x31}, 
one may remove the line between regions 6 and 7 or the line between
regions 5 and 6 to yield $Y^{3,1}$ and $Y^{3,0}$, respectively. 

The Kasteleyn matrix for this tiling is 
\beq
K=\left( \begin{array}{c|cccc}   
                             & \ \ 2 \ \ & \ \ 4 \ \ & \ \ 6 \ \ & \ \ 8 \ \ \\ \hline
                             1 &  1+w^{-1}  &  1  &  0 &  z  \\ 
                             3 & 1 & -1-w & 1 & 0 \\
			     5 & 0 & 1 & w^{-1} & 1 \\
                             7 & z^{-1} & 1 &  1 & -1 \end{array} \right)
\eeq
which has determinant $\det K = 7w^{-1}+w^{-2}+8+w-z^{-1}-z+w^{-1}z$. This yields 
the proper toric diagram and multiplicities for this phase of $X^{3,1}$ \cite{Dunn:2005a}.

\section{Conclusions}
\label{section_conclusions}

In this paper, we have presented a piece of technology which not only
simplifies previously difficult calculations, but also draws new connections to
other interesting areas of study. In this section, we present some conlcuding remarks
about brane tilings and dimers, and suggest some arenas for further study. 

First, we note the computational power of our construction. The correspondence between toric singularities, brane tilings, and dimers presented in the present work goes well beyond any other correspondence yet proposed. 
We establish a precise connection between toric diagrams, GLSM multiplicities, and dimer quantities. This connection enables us to explore toric singularities other than only
the simplest ones, and promises to be extremely useful for the future study of toric
geometries. We further emphasize here that the central object in our construction, the brane 
tiling, is a {\bf physical} object constructed out of NS5-branes and D5-branes in Type IIB string
theory, and not just a mathematical tool. 

Previous to our work, it had been suspected that there was a one-to-one correspondence between toric geometries with specified GLSM multiplicites and toric phases of the quiver gauge theory. The increased computational power provided by the brane tiling has enabled us to show that there is a relatively simple counterexample to this conjecture, that
of Pseudo del Pezzo 5. We find that in fact there are examples where the only way of
distinguishing two phases is by checking graph isomorphism, i.e.~there
exists no simpler characterisation of the graph other than the graph
itself.

The operation of Seiberg duality on a quiver theory is
straightforward, although it 
becomes extremely cumbersome to
implement it many times. The dimer model
allows us to encode Seiberg duality as an operation on the Kasteleyn matrix.  In
many cases this provides an extremely efficient way of enumerating all
toric phases of a quiver theory, as long as one has at least one phase
of the quiver theory to start with. It is an interesting question whether
or not one can explore even non-toric theories via Seiberg duality.

In recent work \cite{Martelli:2005tp}, a geometrical method was
given for performing a-maximization \cite{Intriligator:2003jj} purely in terms of toric data,
without needing to know the explicit Calabi-Yau metric on the toric
space.  This method involved the use of a Monge-Ampere equation to
extremise a certain function.  It is interesting to note that the
solution to a Monge-Ampere equation also appears in the dimer model
literature \cite{Kenyon:2003uj}, where it is related to the
extremization of the surface tension functional that determines the
asymptotic shape (limit curve) of a random matching of many copies of
the fundamental domain of a bipartite graph.  It would be fascinating
to determine whether or not these two extremization problems are related.

In general, there are many different properties of dimer models that have
been studied in the mathematics literature. It is undoubtedly worth
exploring this literature in depth, since there are surely many additional connections
one can draw between the mathematics and physics of dimers. In particular, the
mathematical literature has different examples of partition functions it is possible
to define on a dimer model. Could some of these quantities be related to e.g.~a central charge for the quiver gauge theory, and is it possible to prove
the $a$-theorem for even the simple example of Higgsing?

The limit curve of the dimer model is related to the geometry that is mirror to the toric Calabi-Yau cone.  This mirror geometry encodes the quantum corrections from worldsheet instantons to the classical geometry of the toric CY, and provides a point of contact between dimer models and topological string theory \cite{Okounkov:2003sp}. It would be interesting to investigate this connection in more detail.

One important gap in our understanding of the brane tiling/dimer correspondence is the string theoretic meaning of the perfect matchings.  We proved in section \ref{section_dimers_GLSM} that introducing the dimer model on the bipartite graph of the brane tiling is the correct thing to do to compute the moduli space of the quiver theory, but we do not currently understand the mechanism that produces these perfect matchings in the string theory construction of the brane tiling in terms of 5-branes and strings.

Finally, we note that in this paper, we do not discuss at all the problem of starting with
a given toric geometry and deriving the brane tiling from that. Ideally, one would like to be able to
do this, since there would then be a direct link to the dual SCFT from any given toric
geometry. In fact there has been recent progress on this, and we may say
with confidence that it is possible to derive the brane tiling related to any
toric Calabi-Yau \cite{Vegh:2005}.

{\bf Acknowledgements:} We would like to thank V.~Braun, J.~de Boer, R.~Dijkgraaf, A.~Dunn, Y.-H.~He, P.~Kazakopoulos, T.~Faulkner, A.~Uranga and J.~Walcher for helpful discussions. 


\bibliographystyle{JHEP}

\bibliography{kk.bib}

\end{document}